\documentclass[prb,nobibnotes]{revtex4}
\usepackage{amssymb}
\usepackage{amsmath}
\usepackage{comment}
\usepackage{graphicx}
\usepackage{booktabs}
\usepackage[tracking=true]{microtype}
{\refstepcounter{comment}%
\begin{quote}
\ttfamily\small$\blacksquare$ \textbf{\underline{Comment} $\sharp$\thecomment:}}%
{\end{quote}}

{%\refstepcounter{comment}
\begin{quote}
\ttfamily\small$\blacktriangleright$ \textbf{\underline{Reply} $\sharp$\thecomment:}}%
{\end{quote}}

\newcommand{\xB}{x_{\rm B}}
\newcommand{\Q}{{\cal Q}}

\begin{document}

\author{L\'aszl\'o Jenkovszky}
\affiliation {Bogolyubov ITP, Kiev, 03143 Ukraine; e-mail: jenk\ at\ bitp.kiev.ua}
%\affiliation{BITP, 14b Metrologicheskaya str., 03143 Kiev, Ukraine; jenk@bitp.kiev.ua}
%%%%%%%%%%%%%%%%%%%%%%%%%%%%%%%%%%%%%%%%%%%%%%%%%%%%%%%%%%%%%%%%%%%%%%%%%%

\title{\LARGE{Spin and Polarization in High-Energy Hadron-Hadron and Lepton-Hadron Scattering}}

\begin{abstract}
The role of spin degrees of freedom in high-energy hadron-hadron and lepton-hadron scattering is reviewed with emphasis on the dominant role of soft, diffractive, non-perturbative effects. Explicit models based on analyticity and Regge-pole theory, including the pomeron trajectory (gluon exchange in the $t$ channel) are discussed. We argue that there is a single, universal pomeron in Nature, manifest as relatively ``soft'' or ``hard'', depending on the kinematics considered.
% Regge trajectories are non-linear, complex functions, %predicting a finite number of resonances.
{Both the pomeron and the non-leading (secondary) Regge trajectories, made of quarks are non-linear, complex functions. They are populated by a finite number of resonances: known baryons and mesons in case of the reggeons and hypothetical glueballs in case of the pomeron (“oddballs” on the odderon trajectory).}
Explicit models and fits are presented that may be used in recovering generalized parton distributions from deeply virtual Compton scattering and electoproduction of vector mesons.
\end{abstract}

\keywords{spin; polarization; diffraction; pomeron; gluon; confinement}
%\arxivnumber{}
%%%%%%%%%%%%%%%%%%%%%%%%%%%%%%%%%%%%%%%%%%

%%%%%%%%%%%%%%%%%%%%%%%%%%%%%%%%%%%%%%%%%%%%%%%%%%%%%%%%%%%%%%%%%%%%%%%%%%

\maketitle
%%%%%%%%%%%%%%%%%%%%%%%%%%%%%%%%%%%%%%%%%%%%%%%%%%%%%%%%%%%%%%%%%%%%%%%%%%

%\toccontinuoustrue

%%%%%%%%%%%%%%%%%%%%%%%%%%%%%%%%%%%%%%%%%%%%%%%%%%%%%

\section{Introduction}
Interest in spin physics at high-energies was varying with time. In~general, there is a prejudice that the role of spin decreases as energy is increasing. This may be true in general, especially in hadronic reaction, with~some exceptions. First, one has to understand the origin of polarization in the scattering of unpolarized hadrons, as~for example in $\Lambda$ production. Another issue is the still purely understood delicate mechanism of the diffraction minimum in high-energy proton-proton scattering. This point will be discussed in Section~\ref{Sec:Hadron}. Here one inevitably encounters related problems such as: diffraction and  the nature of the pomeron, { that could be utilized as made of gluons; reconciling soft (non-perturabative, Regge) and hard (perturbative QCD) %Please define if appropriate.
aspects of the theory}, nucleon structure, revealed in inclusive deep-inelastic scattering (DIS) and exclusive deeply virtual Compton  scattering (DVCS) as means to reveal the nucleon structure, closely related to spin and polarization, thus connecting structure and~dynamics.

{The present paper is a recapitulation of the earlier results by the author, revised,  updated and extended with account for the recent developments in the field, especially those connected to “nucleon holography”, based on generalized parton distributions (GPDs). Parallel to these practical developments, the~still open basic problem of quark confinement is always present in our discussions.}

Due to recent progress in understanding the internal structure of the nucleon, especially connected with deeply virtual Compton scattering (DVCS) and generalized parton distributions (GPD) as well as the prospects of building new experimental facilities, such as the planned Electron Ion Collider (EIC), the~interest definitely shifted to lepton induced inelastic~reactions.

Recent interest in spin physics was triggered two decades ago by the publication of the EMC data~\citep{EMC} (so-called ``spin crisis''). { We briefly mention it in Section~\ref{Sec:Crisis} recommending for further reading the excellent presentation of this issue in Ref.~\cite{Rev2}.}

{Since all high-energy processes are dominated by pomeron exchange in the $t$ channel, we  pay special attention to the derivation and properties of the pomeron and its relation to gluon exchange, by~which we mean the identity/diversity of/between the BFKL %Please define if appropriate.
pomeron derived from QCD and that based on the analytic $S$ matrix theory.}   In particular, we discuss the following~issues:
\begin{itemize}
\item Is the input pomeron a simple Regge pole? What are the alternatives, if any?
\item Do resonances terminate, { abruptly replaced by a continuum or they gradually fade, their peaks becoming progressively wider and lower? This issue is connected with the Hagedorn spectrum and possible phase transition between hadronic and quark-gluon matter.}
\item Origin of the diffraction (dip-bump) pattern in elastic hadron scattering;
\item Role of unitarization in producing the dip-bump structure;
\item Are there two (or more) pomerons---a ``soft'' and a ``QCD-inspired'', ``hard'' one?
\item Can the Regge-pomeron pole be $Q^2-$dependent?
\end{itemize}

The paper is organized as follows. In~Section~\ref{Sec:Hadron} we discuss elastic $NN$ %Please define if appropriate.
scattering with spin. Recall~that the dominant point of view is that the role of spin decreases as energy increases, and~it can be ignored at multi-GeV energies, i.e.,~beyond those of the ISR, %Please define if appropriate.
and~even more so at the LHC. %Please define if appropriate.
Spin~effect may be important in the region of the diffraction minimum, i.e.,~around $t=-1$ GeV$^2$. The~point is that the origin and mechanism of the dip-bump phenomenon is still disputable. There are many models but no theoretical understanding of the origin of this important phenomenon. Predictions, mostly based on variants of the Regge-eikonal approach, failed in predicting the position and depth of the dip at the LHC. Spin effects are not the favourite but still a viable possibility, alternative to the dominant ones, based on unitarity (Section \ref{Subs:Unitarity}).

The importance and properties of the Regge trajectories are repeatedly stressed throughout this paper, with~a dedicated Section \ref{Subs:Trajectory}. Dual models have shown that Regge trajectories are sort of dynamical variables. The~trajectories are non-linear, complex function. Their thresholds and asymptotic behaviour should satisfy known constraints. For~example, the~lowest, $4m_{\pi}$ threshold affects the behaviour of the differential cross section in the Coulomb interference region, competing with possible spin~effects.

In Section~\ref{Sec:GPD} we focus on the procedure of recovering Generalized Parton Distributions (GPDs) from electron--proton scattering with spin, measured at the JLab~\cite{Grav}. Here again the pomeron exchange is a basic element of the theory. Different from hadron-hadron scattering, dominated by a ``soft'' pomeron exchange, here the pomeron is ``hard'', i.e.,~its intercept is much larger ($\alpha(0)\approx 1.3)$ than that considered in Section \ref{Sec:Hadron}. In~Section \ref{Subs:Capua} we introduce a model for Deep Inelastic Scattering (DIS) amplitude with $Q^2$-dependent pomeron intercept. That model will be generalized to deeply virtual Compton scattering (DVCS) in Section~\ref{Sec: Reggeomety}, applicable both to ``soft'', e.g.,~$NN$ scattering Section~\ref{Sec:Hadron}, and~``hard''~DVCS.

Various aspects related to the origin of nucleon's spin are discussed in Section~\ref{Sec:Crisis}. We share the point of view that there is no ``crisis'', over-dramatized after the EMS measurements. The~proton spin $1/2$ can be collected from that of its constituents and their~motion.

{Finally, let us mention  recent findings in spin physics revealed in ultra-relativistic heavy-ion collisions; STAR %Please define if appropriate.
collaboration~\cite{STAR} discovered a significantly nonzero global polarization of Lambda  hyperons produced in non-central Au--Au collisions in the RHIC %Please define if appropriate.
Beam Energy Scan (BES) Program~\cite{STAR}. Different hydrodynamic models~\cite{Karp, Karp1, Karp2} generally reproduce the magnitude of the measured polarization. In~the hydrodynamic models, the~Lambda hyperons produced at “particlization” (fluid to particle transition) hypersurface acquire polarization via a thermodynamic spin-vorticity coupling mechanism. This effect is interesting as a possible manifestation of the most vortic fluid ever~made.

Interestingly, this is not the first case that $\Lambda$ polarization raises interest in the high energy community. In~the nineteens of the past century, much discussion was related~\cite{Lambda} to the origin of polarization in the scattering of unpolarized particles.
According to Ref.~\cite{Strum} polarization of inclusively produced lambdas arises due to the spin--orbit interaction in a scalar field in which quarks recombine into hadrons.}

The present paper to a large extent reflects  the author's vision of the spin effects in the ``soft'' region and his personal contribution in the field. For~a much wider panorama we recommend to the reader the excellent papers~\cite{Leader0, Rev1, Rev2}.

\section{Hadron-Hadron~Scattering}\label{Sec:Hadron}

{In this section we collect the basic notions and rules indispensable in describing hadronic reactions. We restrict our discussion of this vast field to elastic nucleon scattering in the nearly forward region, dominated by the diffraction cone, and~nominated “soft” or “non-perturbative”, implying that perturbative QCD methods here are not applicable. Instead, the~basic tools are those based on the analytic $S$ matrix, namely, dispersion relations, unitarity and Regge poles. We relate these means in the context of the subject of the present paper---spin and polarization effects at high energies. Below~is a short overview of the relevant~tools.

Soft events are described by Regge pole models, sometimes appended by “QCD-inspirations”. In~most of the papers on the subject, a~spin-independent invariant scattering amplitude is used at high-energies. Since spin effects cannot be excluded {a priori}, below~we briefly present an introduction to the relevant~formalism.

In the Regge pole model, the~trajectories contain most of the basic information on the dynamics. Their form is constrained by unitarity and analyticity, and~is intimately connected by the Chew--Frautchi plot to the spectrum of hadrons. This is a central issue in high energy physics, since~it is related to the problem of confinement: how do heavy resonances melt producing a quark--gluon soup? In our approach, the~real part of Regge trajectories is limited, implying a finite number of resonances in Nature. This issue is related to the Hagedorn spectrum and “ultimate temperature”, now interpreted as a the temperature of the phase transition between hadrons and quark--gluon plasma~\cite{Biro}.

Unitarity is both important and technically difficult to be satisfied. Various approaches and approximations to unitarity are presented and compared in this Section. Let us mention that the final result (construction of a viable scattering amplitude) depends both on the input and subsequent unitarization procedure. The~better the input (“born term”), the~better are chances for its subsequent successful unitarization: unitarity correction to a reasonable input (for example, a~dipole pomeron, reproducing itself under unitarization) are small~\cite{DP}.}

\subsection{Elastic Proton-Proton~Scattering}

Elastic proton--proton scattering is described by
five helicity amplitudes~\cite{Lehar, Lehar1}, that
in the  Regge limit, $t$ fixed and $s \rightarrow \infty$,
can be expressed in terms of the Regge-pole contributions as~\cite{Capella, Troshin, Troshin1, Selyugin1, Selyugin2}
\begin{eqnarray}
\Phi_{\lambda_1, \lambda_2,\lambda_3, \lambda_{4}} (s,t) \approx
\sum_{i} g^{i}_{\lambda_1, \lambda_2}(t)  g^{i}_{\lambda_3, \lambda_4}(t)
[\sqrt{|t|}]^{|\lambda_1- \lambda_2|+|\lambda_3- \lambda_4|}
\left(\frac{s}{s_0}\right)^{\alpha_{i}} (1 \pm e^{-i \pi \alpha_{i} }),
\end{eqnarray}
where $\alpha_{i}\equiv\alpha_{i}(t)$ are Regge trajectories. In~high-energy elastic scattering the dominant trajectory is that of the pomeron. In~the simplest case, it is linear, parametrized, for~simplicity as $\alpha(t)=\alpha(0)+\alpha' t$. We will use also advanced models of Regge trajectories, to~be introduced in the next Section \ref{Subs:Trajectory}. Recently~\cite{Glueballs} it was used to predict glue and oddballs---resonances made of two or three~gluons.

The above helicity amplitudes differ only by powers of $\sqrt{t}$ leaving the Regge propagator (i.e.,~nature of the Regge pole) intact. This means that the addition of spin-flip amplitudes by themselves will not provide the dip mechanism, although~they will slightly modify it. The~(single) diffraction minimum followed by maximum was shown~\cite{DP} to arise even at the Born level by choosing the pomeron to be a dipole  (double Regge pole or dipole pomeron (DP) \cite{DP}---an interesting and unique alternative to a simple Regge pole). Fits to the the data on $pp$ scattering, including those from the LHC (TOTEM Collaboration) using DP %Please define if appropriate.
was performed recently~\cite{DP1}.
The global fit to all energies, including the dip-bump is good, although~it leaves room for further perfection. Inclusion of the spin-flip component may improve the~situation.

In most of the relevant papers the dip-bump structure is generated by unitarity corrections, see~Section \ref{Subs:Unitarity}, although~there is no consensus on the~details.

The differential cross sections is given by
\begin{eqnarray}
\frac{d \sigma}{dt} = \frac{2 \pi}{s^2} \left( |\Phi_{1}|^2 + |\Phi_{2}|^2+
|\Phi_{3}|^2 + |\Phi_{4}|^2+4 |\Phi_{5}|^2\right)\,.
\end{eqnarray}

The helicity amplitudes can be written as $\Phi_{i}(s,t) =
\Phi^{h}_{i}(s,t)+\Phi^{\rm em}_{i}(s,t) e^{\varphi(s,t)}$\,, where
$\Phi^{h}_{i}(s,t)$ comes from the strong interactions,
$\Phi^{\rm em}_{i}(s,t)$ from the electromagnetic interactions and
$\varphi(s,t)$
is their~interference.

The spin correlation parameters, the~analyzing power---$A_N$ and the
and double-spin parameter $A_{NN}$, can be extracted from experimental measurements:
\begin{eqnarray}
A_{N} &=&\frac{ \sigma(\uparrow)- \sigma(\downarrow)}{\sigma(\uparrow)+ \sigma(\downarrow)} =
\frac{\Delta \sigma^{s}}{\sigma_{0}}\,,\\
A_{NN} &=&\frac{ \sigma(\uparrow \uparrow) - \sigma(\uparrow \downarrow)}{\sigma(\uparrow \uparrow)+
\sigma(\uparrow \downarrow) } = \frac{\Delta \sigma^{d}}{\sigma_{0}}\,,
\end{eqnarray}
where $\Delta \sigma^{s}$ and $\Delta \sigma^{d}$ refer to the difference of
single- and double-spin-flip cross sections.
The expressions for these parameters are
\begin{eqnarray}
A_{N} \frac{d \sigma}{dt} &=& -\frac{4 \pi}{s^2}
\left [ Im (\Phi_{1} + \Phi_{2}+ \Phi_{3} - \Phi_{4}) \Phi^{*}_{5}\right ]\,;\\
A_{NN} \frac{d \sigma}{dt} &=& \frac{4 \pi}{s^2}
\left [ Re (\Phi_{1} \Phi^{*}_{2} - \Phi_{3} \Phi^{*}_{4}) + |\Phi_{5}|^2\right ]\,.
\end{eqnarray}

\subsection{Regge~Trajectories}\label{Subs:Trajectory}

\begin{figure}[h]
\centering
\includegraphics[width=.99\textwidth]{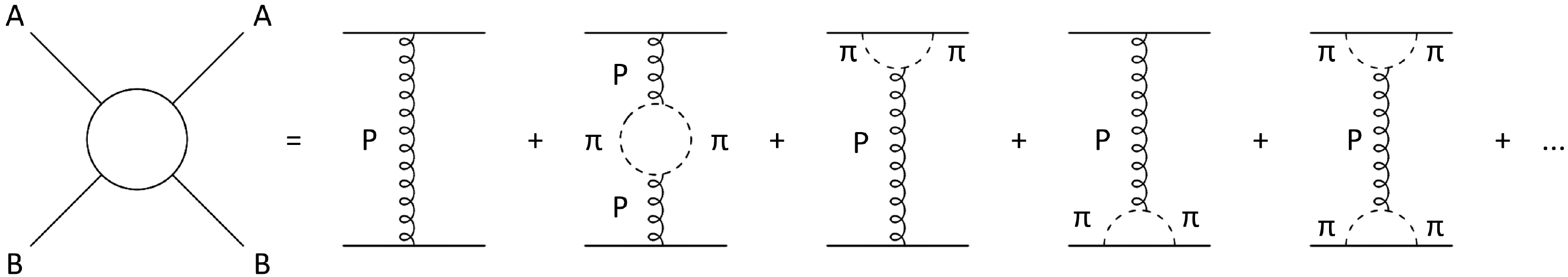}
\caption{Diagram of elastic scattering with $t$-channel exchange containing a branch point at $t=4m_{\pi}^2$.}
\label{Fig:Diagram1}
\end{figure}

Unitarity imposes a severe constraint on the threshold behaviour of the trajectories, as~shown on the diagram, Figure~\ref{Fig:Diagram1}:
\begin{equation} \label{thr}
\Im m\alpha(t)_{t \rightarrow t_0}\sim(t-t_0)^{\Re e\alpha(t_0)+1/2},
\end{equation}
while asymptotically the trajectories are constrained by
\begin{equation}\label{asympt}
\Bigg\vert\frac{\alpha(t)}{\sqrt{t}\ln{t}}\Bigg\vert_{t\rightarrow\infty}\leq {\rm const}.
\end{equation}

The above asymptotic constraint can be still lowered to a logarithm by imposing wide-angle power behaviour for the~amplitude.

The trajectory satisfying the above constraints is:
\begin{equation}\label{3}
\alpha(t) = \frac{1+\delta+\alpha_{1}t}{1+\alpha_{2}\big(\sqrt{t_{0}-t}-\sqrt{t_0})},
\end{equation}
where $t_0=4m_{\pi}^2$ for the pomeron and $t_0=9m_\pi^2$ for the odderon and $\delta, \alpha_1, \alpha_2$ are parameters fitted to the data with the obvious constraints: $\alpha(0)\approx 1.08$ and
$\alpha'(0)\approx 0.3$ (in the case of the pomeron trajectory). Trajectory (\ref{3}) has square-root asymptotic behaviour, in~accordance with the requirements of the analytic $S$-matrix~theory.

\begin{figure}[h]
\centering
\includegraphics[width=1\textwidth]{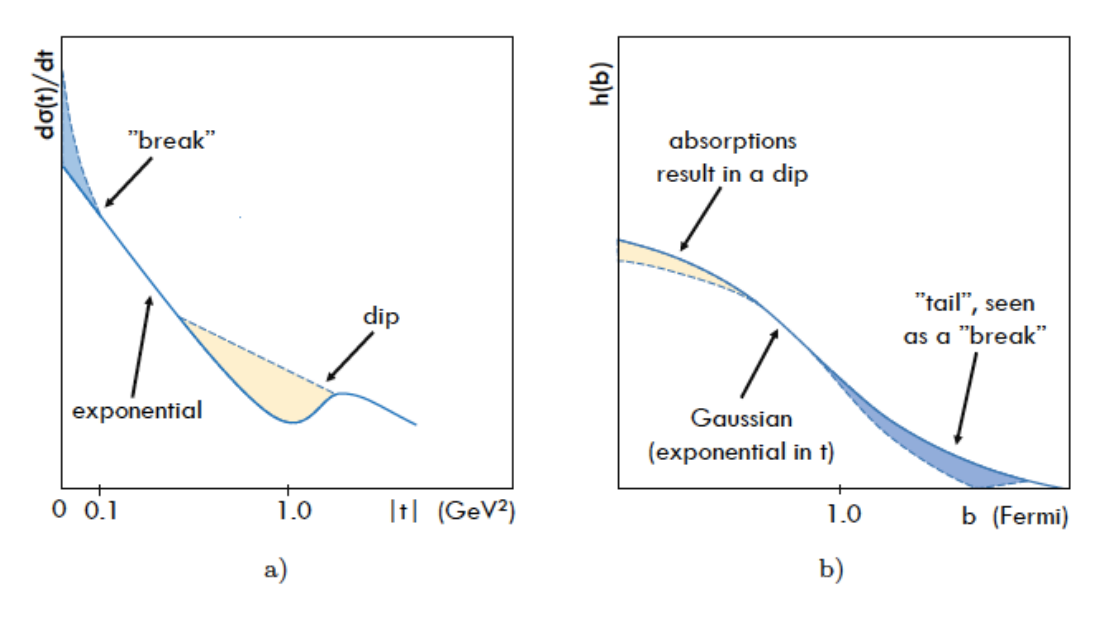}
\caption{(\textbf{a}) Typical shape of the high-energy differential (diffractive) cross section, and
(\textbf{b}) Its impact-parameter image. A ``dictionary'' relating the shape of the diffraction cone (in $t$), left icon to its impact-paramter image (rignt), and~highlighting their two silent features: the ``break'', origin of nucleon's atmosphere and the dip-bump, resulting from absorptions (unitarity!) at small values of the impact {parameter, see}~\cite{DP}.}
%mdpi: Please confirm whether the references are correct, There is no DP2 find in references.
\label{Fig:Diagram}
\end{figure}

{The contribution of the lowest threshold in the $t$ channel of the trajectory and the scattering amplitude is shown on the diagram, Figure~\ref{Fig:Diagram1}. The~presence of this threshold produces a ``break'' in the diffraction cone and generates the pion cloud of the nucleon, as~shown in Figure~\ref{Fig:Diagram}.}

For $t>>t_0,\ \  |\alpha(t)|\rightarrow\frac{\alpha_1}{\alpha_2}\sqrt{|t|}$. For~$t>t_0$ (on the upper edge of the cut), $\Im m\alpha>0.$

The intercept is $\alpha(0)=1+\delta$ and the slope at $t=0$ is
\begin{equation}\label{Slope}
\alpha'(0)=\alpha_1+\alpha_2\frac{1+\delta}{2\sqrt{t_0}}.
\end{equation}

Resonances terminate, {i.e.,~they do not appear beyond} the maximum value or the {finite} real part of a given Regge~trajectory.

For a recent application of this trajectory to glueball spectroscopy see Ref.~\cite{Glueballs}.

\subsection{Unitarity}\label{Subs:Unitarity}

{The best known unitarization procedure is that of Regge-eikonal, where the input (``Born term'') is a Regge-pole~amplitude.

It implies two Fourier--Bessel transforms (integrals), one from the Mandelstam variable $t$ (momentum space) to  the impact parameter $b$ (coordinate space). While direct integration is usually performed analytically (trivial for linear Regge trajectories and still simple for a single square-root threshold, see Ref.~\cite{NC}, the~inverse integral is much more cumbersome, and~can be done only numerically. The~technicalities become even more complicated when spin degrees are included (see below).}

A simple unitarization procedures is by the use~\cite{Selyugin1, Selyugin2} of the equation:
\begin{eqnarray}
\frac{dN}{dy} =\Delta N \ [ \ 1 - N \ ]\,, \label{log-eq}
\end{eqnarray}
where $y =\log(s/s_0)$ and $\Delta = 1- \alpha(0)$, ($\alpha(0)$
is the intercept of the leading Regge pole~trajectory.

Its solution is
\begin{eqnarray}
N = \frac{\chi(s,b)}{1 \ + \ \chi(s,b)}\,,
\end{eqnarray}
where $\chi(s,b)\approx s^{\Delta}$ is connected with the Born term of the scattering~amplitude.

The scattering amplitude is
\begin{eqnarray}
\Phi^{h}(s,t) \ = \  \frac{i}{2  \pi}
\ \int \ d^2 b  \  e^{i \vec{b} \vec{q} } \frac{\chi(s,b)}{1+ \chi(s,b)}\,.
\label{K-mat}
\end{eqnarray}

The  phase $\chi(s,b)$  is connected to  the interaction
potential:
\begin{eqnarray}
\chi(s,b)\ = \ F_{\rm Born}(s,b)  \ \approx   \frac{1}{k}
\ \int  \hat{V}\left( \sqrt{b^2  + z^2}  \right) dz.
\label{potential3}
\end{eqnarray}

If the potential contains a non-spin-flip part and, for~example,
spin-orbital and spin-spin interactions, the~phase is:
\begin{eqnarray}
\chi(s,b) = \chi_{0}(s,b)
- i \ \vec{n} \cdot (\vec{\sigma}_{1} + \vec{\sigma}_{2} ) \chi_{\rm LS}(s,b)
- i  (\vec{\sigma}_{1} \cdot \vec{\sigma}_{2} ) \ \chi_{\rm SS}(s,b).
\end{eqnarray}

If however only the spin-flip and spin-non-flip parts are accounted for, the~overlap function is
\begin{eqnarray}
\Gamma(s,b)=\frac{\chi_{0}(s,b) +\sigma \chi_{sf}(s,b)}{1
+\chi_{0}(s,b) +\sigma \chi_{sf}(s,b)}= 1- \frac{\left(1+\chi_{0}(s,b)\right) -
\sigma \chi_{sf}(s,b) }{\left(1 + \chi_{0}(s,b)\right)^2 - \left(\sigma \chi_{sf}(s,b)\right)^2}\,.
\label{K-m-s}
\end{eqnarray}

Using the representation for the Bessel functions
\begin{eqnarray}
J_{0}(x) \ =  \frac{1}{2 \pi}
\ \int_{0}^{2 \pi} \  e^{i x \cos \phi }
\ d \phi \ \ \ \ \
J_{1}(x) \ =  -\frac{1}{2 \pi}
\ \int_{0}^{2 \pi} \  e^{i x \cos \phi } \  \sin{\phi}
\ d\phi\,,
\label{J0}
\end{eqnarray}
the representation of spin-non-flip and spin flip amplitude becomes
\begin{eqnarray}
\Phi^{h}_{1}(s,t) \ = \ i
\ \int_{0}^{\infty} \ b J_{0}(b q) \frac{\chi_{0}(s,b)}{1+\chi_{0}(s,b)} db \,;
\label{K-matrix}
\end{eqnarray}
\begin{eqnarray}
\Phi^{h}_{5}(s,t) \ = \ i
\ \int_{0}^{\infty} \ b^2 J_{1}(b q) \frac{\chi_{sf}(s,b)}{\left(1+\chi_{0}(s,b\right))^2} db \,.
\label{K-m-spin}
\end{eqnarray}

\subsubsection{``$U$-matrix''~Unitarization}
With an extra coefficient $n$
\begin{eqnarray}
\frac{dN}{dy} =\Delta N \ [ \ 1 - N/n \ ]\,. \label{UT-eq}
\end{eqnarray}
one gets the so-called $U$-matrix unitarization form, see~\cite{Troshin} and references~therein.

{This way of unitarization, developed mainly in Serpukhov, Dubna and Kiev, is less familiar than the eikonal one, nevertheless it results in a number of interesting predictions, among~which is the so-called reflective scattering~\cite{Troshin2}, resulting in an increasing ratio
$\sigma_{el}(s)/\sigma_{tot}(s)$ at the LHC. The~unorthodox predictions of the $U$ matrix approach may bringing new ideas in the complicated field of ``soft physics''.}

In the impact parameter representation, the~properties of the $U$-matrix were  explored in
Ref.~\cite{Troshin}. In~that approach, the~hadronic amplitude is given by
\begin{eqnarray}
\Phi^{h}(s,t) \ = \ \frac{i}{2  \pi}
\ \int \ d^2b  \  e^{i \vec{b} \vec{q} } \frac{\chi(s,b)}{1+ \chi(s,b)/2}\,,
\label{U-mat}
\end{eqnarray}
where $\chi(s,b)$ is the same Born amplitude as~before.

Comparing Equation~(\ref{U-mat}) with Equation~(\ref{K-mat}), we see that they are of similar
rational form, just they differ by the additional coefficient in the denominator.
This additional coefficient leads to different analytic properties: the upper bound
at which the overlapping function saturates
will be twice that compared with the eikonal or the $U_e$ representations,
and the inelastic
overlap function at $b=0$ tends to zero at high energies. Furthermore, $\sigma_{\rm el}/\sigma_{\rm tot} \rightarrow 1$.

For the $pp$ helicity amplitudes
the solution of the unitarity equations is~\cite{Troshin}:
\begin{eqnarray}
\Phi_{\lambda_3,\lambda_4,\lambda_1,\lambda_2}({\bf p},{\bf q}) & =
& U_{\lambda_3,\lambda_4,\lambda_1,\lambda_2}({\bf p},{\bf q})+ \label{heq}\\
& & i\frac{\pi}{8}
\sum_{\lambda ',\lambda ''}\int d\Omega_{{ \bf \hat k}}
U_{\lambda_3,\lambda_4,\lambda ',\lambda ''}({\bf p},{\bf k})
\Phi_{\lambda ',\lambda '',\lambda_1,\lambda_2}({\bf k},{\bf q})\,,\nonumber
\end{eqnarray}
%in the impact parameter representation

\subsubsection{Eikonal}
To get the standard eikonal representation of the elastic scattering amplitude
in the impact parameter representation one uses the non-linear equation
\begin{eqnarray}
\frac{dN_e}{dy} = -\Delta \log(1-N_e) [1-N_e]\,. \label{nl-eik}
\end{eqnarray}
where $y =\log(s/s_0)$ and the subscript ``e`` implies that the solution $N_e$
is of the standard eikonal form
\begin{eqnarray}
N_e = \Gamma (s,b)= [1- e^{- \chi(s,b)}]. \label{nl-eik1b}
\end{eqnarray}

The eikonal representation is then
% neglecting the   spin-spin correlation part of the interaction potential, is
\begin{eqnarray}
\Phi^{h}(s,t) \ =  \frac{i}{2  \pi}
\ \int \  e^{i \vec{b} \vec{q} } \ \left[1 - e^{- \chi(s,b) }\right]
\ d^2 b,
\label{tot0}
\end{eqnarray}
where:
\begin{eqnarray}
\chi(s,b) = \chi_{0}(s,b)
- i \ \vec{n} \cdot (\vec{\sigma}_{1} + \vec{\sigma}_{2} ) \chi_{LS}(s,b)
- i  (\vec{\sigma}_{1} \cdot \vec{\sigma}_{2} ) \ \chi_{SS}(s,b).
\end{eqnarray}

With account for Equations~(\ref{J0}) and (\ref{J1}), one gets respectively for spin non-flip and spin-flip
\begin{eqnarray}
\Phi%%%%%%%%
^{h}_{1}(s,t) \ =  \  i
\ \int_{0}^{\infty} \ b J_{0}(b q)\left[ 1- e^{\chi_{0}(s,b)}\right] \
\left[1 - b^2 \chi^2_{\rm LS}(s,b) -3/2 \chi^2_{\rm SS}(s,b)\right]
\ d b,
\label{J1}
\end{eqnarray}
\begin{eqnarray}
\Phi^{h}_{5}(s,t) \ =   \ i
\ \int_{0}^{\infty} \  J_{1}(b q) \chi_{1} e^{\chi_{0}(s,b)}
\ b \  \left[\chi_{\rm LS}(s,b) +i \ \chi_{\rm LS}(s,b) \ \chi_{\rm SS}(s,b)\right]  \ d b,
\label{tot0a}
\end{eqnarray}
where
\begin{eqnarray}
\chi(s,b)_{0}  &\approx &
\ \int_{-\infty}^{\infty}  V_{0}(s,b,z)   dz\,; \\
\chi(s,b)_{1}  &\approx  & \frac{b}{2}
\ \int_{-\infty}^{\infty}  V_{1}(s,b,z)   dz\,.
\label{potential5}
\end{eqnarray}

For Gaussian potentials $V_{0}$ and $V_{1}$
\begin{eqnarray}
V(s,b)_{0,1} \ \approx
\ \int_{-\infty}^{\infty} e^{-B r^2}  \ dz = \frac{\sqrt{\pi}}{\sqrt{B}} e^{-B b^2}\,,
\label{potential2}
\end{eqnarray}
in the first Born approximation, $\Phi_{0}^{h}$ and $\Phi_{1}^{h}$ take the forms
\begin{eqnarray}
\Phi^{h}_{1}(s,t) \ \approx
\ \int_{0}^{\infty} \ b J_{0}(b q) e^{-B b^2} d b = e^{-B q^2};
\label{tot0b}
\end{eqnarray}
\begin{eqnarray}
\Phi^{h}_{5}(s,t) \ \approx
\ \int_{0}^{\infty} \ b^2 J_{1}(b q) e^{-B b^2} d b =  q  B e^{-B q^2}.
\label{tot01}
\end{eqnarray}

{In view of these technical complications, of~interest is the unique case of a dipole pomeron (DP), reproducing itself under eikonal corrections, as~noticed in Ref.~\cite{Phillips}, see also the Appendix in~\cite{Fortschritte}. In~other words, DP in~the impact parameters representation remains stable (intact) under eikonalization. In addition, a~DP produces logarithmically rising cross sections at unit pomeron intercept and obeys geometrical scaling, making it attractive as an efficient model of diffraction. Note that higher order Regge poles violate unitarity}.

%%%%%%%%%%%%%%%%%%%
%\newpage
\section{Nuclear Structure: From Deeply Virtual Compton Scattering (DVCS) to Generalized Parton Distributions (GPDs)}\label{Sec:GPD}
%%%%%%%%%%%%%%%%%%%%%%%%%%%%%%%%%%%%%%%%%%%%%%%%%%%%%%%%%%%%%%%%%%
An important part of spin physics is lepton-hadron scattering.
Deeply virtual exclusive $ep$ processes provide an important
tool in accessing the generalized  parton distributions \mbox{(GPDs)
\cite{Mueller:1998fv,Radyushkin:1996nd,Ji:1996nm}.}
The underlying mechanism is Regge-pole exchange in the $t$-channel.
Based on factorization theorems~\cite{Collins:1996fb,Collins:1998be}, GPDs offer
a partonic interpretation of these processes,
where unobserved transverse degrees of freedom are integrated out.
Thereby, these universal functions, defined in terms of matrix elements of quark
and gluon operators or, alternatively,
as a non-diagonal overlap of light-cone wave functions
~\cite{Diehl:1998kh,Diehl:2000xz},
encode the non-perturbative aspects of the nucleon~\cite{Diehl:2003ny,Belitsky:2005qn}.
In particular, GPDs~provide access to the transverse spatial distribution of
patrons~\cite{Burkardt:2000za,Die02,KogSop70}, and~to the decomposition of the nucleon spin in
terms of quark and gluon degrees of freedom~\cite{Ji:1996ek}.

\begin{figure}[h]
\centering
\includegraphics[width=0.94\textwidth]{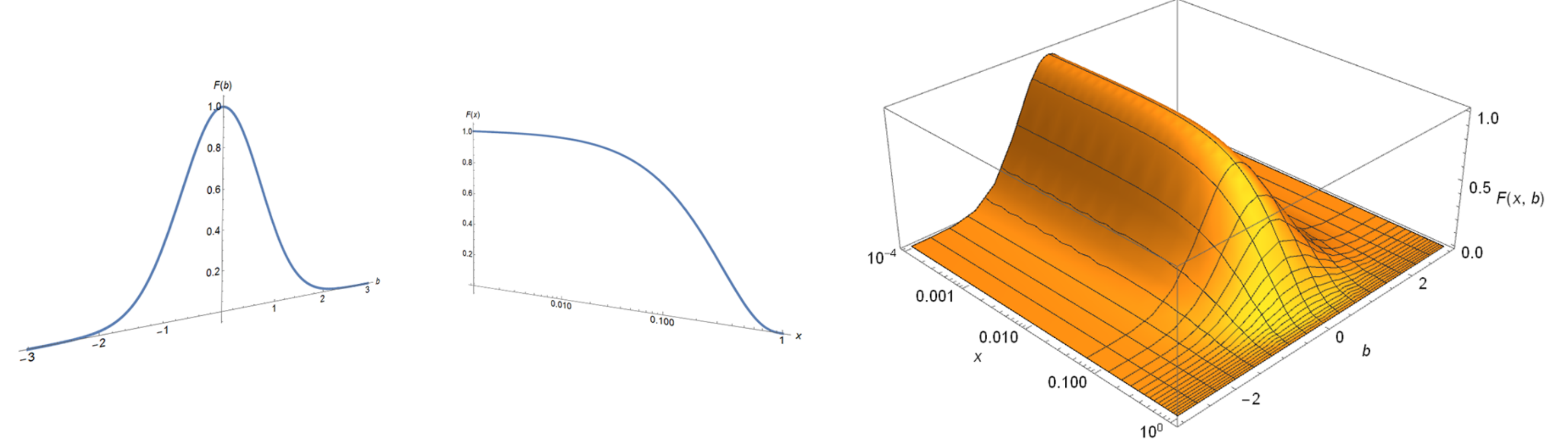}
\caption{Generalized parton distributions ({GPD}) (\textbf{right}) unify in a non-trivial way the the impact parameter image of a nucleon (\textbf{left}) with the $x$ dependence of DIS structure functions (\textbf{middle}).}
\label{Fig:Holography}
\end{figure}

Phenomenologically, exclusive electroproduction of a real photon, DVCS, depicted in Figure~\ref{fig:DVCS-BH} (left),
is the golden channel to constrain GPDs as it is theoretically clean and
the phase of its amplitude can be measured using the interference with
the Bethe--Heitler (BH) amplitude
(see Figure~\ref{fig:DVCS-BH}; \mbox{cf.~Figures~\ref{Fig:Denes} and \ref{Fig:Holography})}.

More details, especially those related to experiments can be found in~\cite{Accardi:2012hwp, Fazio}.

%%%%%%%%%%%%%%%%%%%%%%%%%%%%%%%%%%%%%%%%%%%%%%%%%%%%%%%%%%%%%%%%%%%%%%%%%%
\subsection{Deeply Virtual Compton~Scattering}
\label{sec:presendDVCS}
%%%%%%%%%%%%%%%%%%%%%%%%%%%%%%%%%%%%%%%%%%%%%%%%%%%%%%%%%%%%%%%%%%%%%%%%%%

\begin{figure}[h]
\centering
\includegraphics[width=0.85\textwidth]{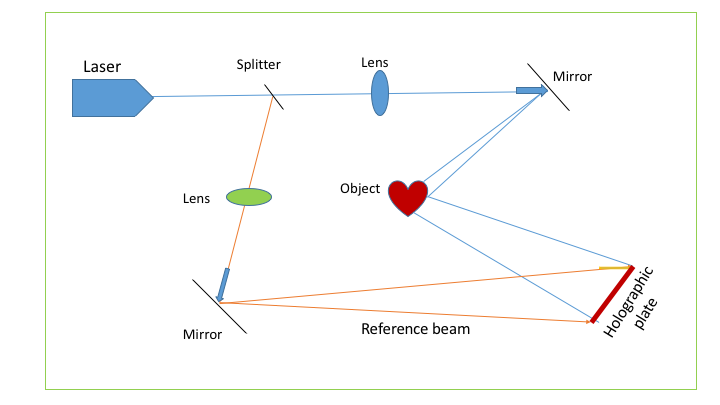}
\caption{{Interference in} DVCS (Equation (\ref{X-electroproduction})) provides a holographic picture of the proton similar to the classical~one.}
\label{Fig:Denes}
\end{figure}

The differential photon electroproduction cross section
is the sum of the BH amplitude squared, DVCS amplitude squared,
and the interference (INT) terms (see Figure~\ref{Fig:Denes})
\begin{equation}
\label{X-electroproduction}
\frac{d\sigma^{ep\to ep\gamma}}{d\xB dt d\Q^2  d\phi d\varphi} =
\frac{d\sigma^{ep\to ep\gamma,\mbox{\tiny BH}}(F_1,F_2)}{d\xB  dt d\Q^2  d\phi d\varphi} \pm
\frac{d\sigma^{ep\to ep\gamma,\mbox{\tiny INT}}(F_1,F_2,{\cal F})}{d\xB dt d\Q^2 d\phi d\varphi} +
\frac{d\sigma^{ep\to ep\gamma,\mbox{\tiny (D)VCS}}({\cal F},{\cal F}^\ast)}{d\xB dt  d\Q^2 d\phi d\varphi}\,.
\end{equation}

\begin{figure}[h]
\begin{center}
\includegraphics[width=0.8\textwidth]{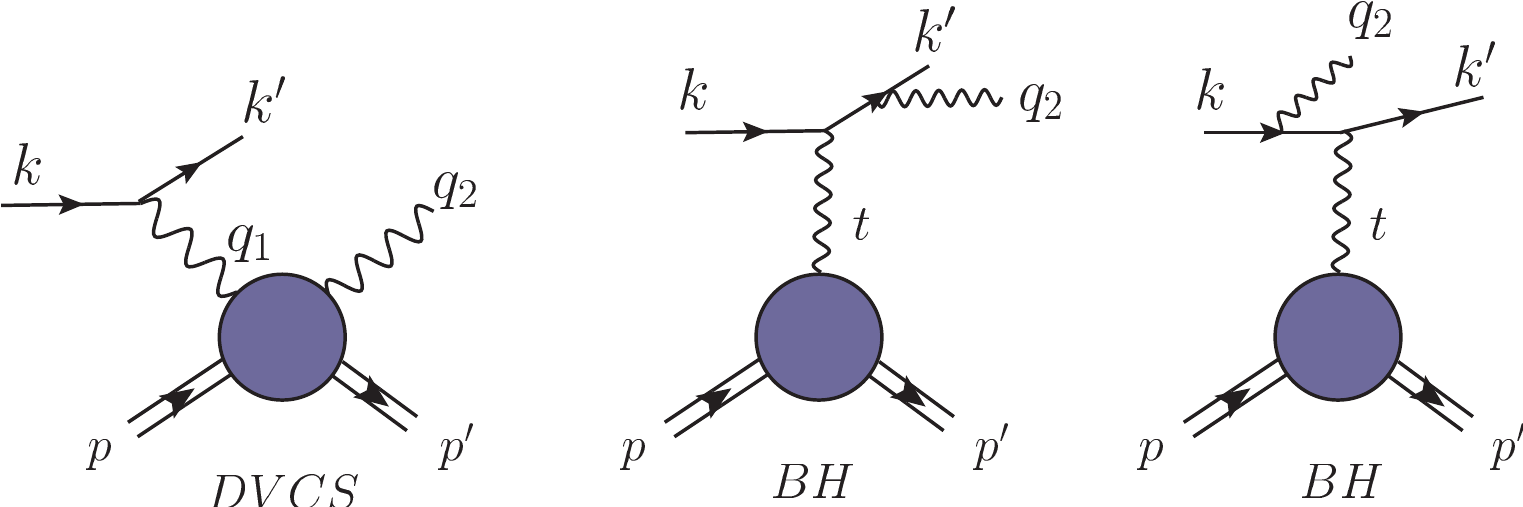}
\end{center}
\caption{
\label{fig:DVCS-BH}
\small Deeply virtual Compton scattering (DVCS) and Bethe--Heitler (BH) amplitudes contributing to the photon leptoproduction
cross section in leading order approximation of QED.} %Please define if appropriate.
\end{figure}
%%%%%%%%%%%%%%%%%%%%%%%%%%%%%%%%%%%%%%%%%%%%%%%%%%%%%%%%%%%%%%%%%%%%%

Here the $+ (-)$ signs
correspond to the electron (positron) beam,
$\xB$ is the common Bjorken scaling variable,
$\phi$ is the azimuthal angle between lepton and hadron scattering planes, and~$\varphi=\Phi-\phi$, where $\Phi (\equiv\phi_S)$
is the angle between the lepton scattering plane and a possible transverse spin
component
of the incoming proton at rest.
To the leading order (LO) in the electromagnetic
fine structure constant $\alpha_{\rm em} = \frac{e^2}{4\pi} \approx
\frac{1}{137},$ and neglecting the electron mass, the~three terms on the
r.h.s.~of (\ref{X-electroproduction}) are known in terms of
the electromagnetic Pauli form factor $F_1(t)$ and the Dirac form factor $F_2(t)$,
parameterizing the BH amplitude, and~a set of 12 photon helicity dependent
CFFs %Please define if appropriate.
${\cal F}_{a b}(\xB,t,\Q^2)$, parameterizing the DVCS amplitude, see
Figure~\ref{fig:DVCS-BH}. These CFFs are labelled by the helicities of the incoming
$a \in \{+,0,-\}$ and outgoing photon $b\in \{+,-\}$
called
\begin{equation}
\label{cffF_{ab}}
{\cal F}_{a b} \in \{{\cal H}_{ab},{\cal E}_{ab}, \widetilde{\cal H}_{ab},
\widetilde{\cal E}_{ab} \} \quad\mbox{with}\quad
{\cal F}_{0 -} = {\cal F}_{0 +}\,, \quad  {\cal F}_{+ -}=  {\cal F}_{- +}\,,
\end{equation}

\unskip

%%%%%%%%%%%%%%%%%%%%%%%%%%%%%%%%%%%%%%%%%%%%%%%%%%%%%%%%%%%%%%%%%%%%%%%%%%%%%%%%%%%%%%%%
\subsection{Relating DVCS Observables to~GPDs}
\label{sec:GPD22CFFs}
%%%%%%%%%%%%%%%%%%%%%%%%%%%%%%%%%%%%%%%%%%%%%%%%%%%%%%%%%%%%%%%%%%%%%
%%%%%%%%%%%%%%%%%%%%%%%%%%%%%%%%%%%%%%%%%%%%%%%%%%%%%%%%%%%%%%%%%%%%%

%%%%%%%%%%%%%%%%%%%%%%%%%%%%%%%%%%%%%%%%%%%%%%%%%%%%%%%%%%%%%%%%%%%%%
GPDs, denoted as
$$
F(x,\eta=\xi, t, \mu^2) \quad\mbox{with}\quad F\in\{H,E,\widetilde{H},\widetilde{E}\}\,,
$$
besides depending on the partonic momentum fraction $x$
and the momentum transfer squared $t$, depend
also on the $t$-channel longitudinal momentum fraction $\eta$, called {skewness}
(often denoted by $\xi$ in the literature),
and on the factorization scale $\mu^2$.
The unpolarized parton GPDs are called $H$ and $E$ \cite{Ji:1998pc}, where the former (latter)
GPD can be loosely associated with a proton helicity (non)conserved~distribution.

DVCS observables can be evaluated in terms of the helicity CFFs (\ref{cffF_{ab}}).
To express them in terms of GPDs, it is
appropriate to utilize a conventionally defined GPD-inspired
CFF basis, such as the one introduced in~\cite{Belitsky:2001ns}:
\begin{equation}
\label{cffF}
{\cal F} \in \{{\cal H},{\cal E}, \widetilde{\cal H}, \widetilde{\cal E}, {\cal H}_3,{\cal E}_3,
\widetilde{\cal H}_3, \widetilde{\cal E}_3,
{\cal H}_{\rm T},{\cal E}_{\rm T}, \widetilde{\cal H}_{\rm T}, \widetilde{\cal E}_{\rm T} \}\,.
\end{equation}

Here, the~CFFs ${\cal H},{\cal E}, \widetilde{\cal H}$, and~$\widetilde{\cal E}$ are
associated with twist-two GPDs $F \in \{H,E,\widetilde{H}, \widetilde{E}\}$ and govern
the photon helicity non-flip DVCS amplitude, i.e.,~at leading twist-two accuracy one has \begin{eqnarray}
{\cal F}_{++}(\xB,t,\Q^2) = {\cal F}(\xB,t,\Q^2) + {\cal O}(1/\Q^2)\quad\mbox{for}\quad{\cal F}
\in \{{\cal H},{\cal E}, \widetilde{\cal H}, \widetilde{\cal E}\}.
\end{eqnarray}

\begin{figure}[h]
\begin{center}
\includegraphics[width=0.95\textwidth]{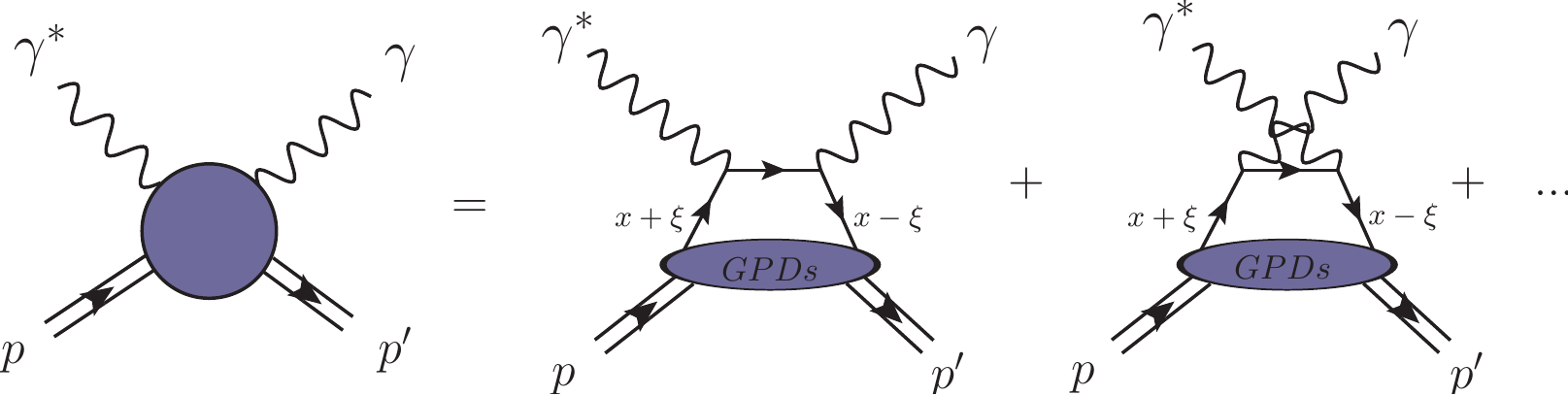}
\end{center}
\caption{
\small
Factorization of the DVCS amplitude to leading order in perturbative QCD and to leading
twist-two accuracy.
\label{fig:DVCS2GPD} }
\end{figure}

To LO they are calculated from the  handbag diagram, depicted in Figure~\ref{fig:DVCS2GPD}, yielding
the convolution formula
\begin{equation}
\label{cffs}
{\cal F}(\xB,t,\Q^2)\stackrel{\rm LO}{=}\sum\limits_{i}\int_{-1}^1 \! dx \left[ \frac{e^2_i}{\xi-x-i\epsilon} \mp \{ x \rightarrow -x \} \right] F_i(x,\xi,t,\mu^2)\;\; \mbox{for}\;\;
{\cal F} \in \left\{ {{\cal H},{\cal E}  \atop \widetilde{\cal H}, \widetilde{\cal E}}
\right\},
\end{equation}
where $e_i$  are the fractional quark charges. The~variable
$\xi\sim \xB/(2-\xB)$ is a conventionally defined Bjorken-like scaling variable, equated to
the longitudinal momentum fraction in the $t$-channel,
and~\mbox{$\mu^2 \sim \Q^2$} being the factorization~scale.

For twist-two and LO accuracy, and~light quarks one adopts the conventions
\begin{eqnarray}
\xi= \frac{\xB}{2-\xB} \quad\mbox{and} \quad \mu^2= \Q^2\,.
\end{eqnarray}

From Regge-pole models, consistent with the data, see Section~\ref{Sec:Models} below, the~real part of the dominant CFF ${\cal H}$ in the small- and even
moderate-$\xB$ region
is much smaller than its imaginary part (at least for smaller values of $-t$). One concludes that the
interference term is negligible and one can simplify the $t$-differential cross section to
\begin{eqnarray}
\label{XTOT-EIC}
\frac{d\sigma^{\mbox{\tiny TOT}}}{dt} \approx
\frac{y^2  \left[ \frac{d\sigma_{\rm T}^{\mbox{\tiny BH,red}}}{dt}  + \varepsilon(y) \frac{d\sigma_{\rm L}^{\mbox{\tiny BH,red}}}{dt} \right] }{\left(1-y \frac{(1-\xB) t}{\Q^2+t} \right) \left(\frac{\Q^2+t}{\Q^2+\xB t}-y\right)}
+ \frac{d\sigma^{\mbox{\tiny\rm DVCS}}(y)}{dt}
\end{eqnarray}
with $\frac{d\sigma^{DVCS}(y)}{dt} =
\frac{d\sigma_{T}^{DVCS}}{dt} + \varepsilon(y)\frac{d\sigma_{L}^{DVCS}}{dt}.$

\section{Modelling~DVCS}\label{Sec:Models}

Below we present models of DVCS and vector meson production (VMP), potentially useful in calculating~GPDs.

Regge pole models provide an adequate framework to describe high-energy, low $t$ scattering phenomena. Being part of the $S$ matrix theory, however,
strictly speaking, they are valid only for the scattering of on-mass-shall particles. Still, the~successful application of the Regge pole models in
describing the off-mass-shall HERA data opened the way to their use in deep-inelastic scattering, DVCS  and vector meson production at HERA. Off-shell extension was realized e.g.,~by calling the $Q^2$-dependent Regge trajectories ``effective'' ones. A~particularly simple and efficient
Regge pole model~\cite{Capua} with $Q^2$-dependent residues (vertices) is presented below, in~Section \ref{Subs:Capua}.

In a more advanced Regge-pole model, {Section} \ref{Sec: Reggeomety} the Regge trajectories and the residues do not depend on virtuality. Instead, the~amplitude contains two (or more) Regge-pole terms, whose relative weight depends on $Q^2$, mimicking the multi-pole nature of the so-called QCD~pomeron.
%mdpi: There is no {Subs: Reggeomety} find in the page, please add.

\subsection*{Simple Model of~DVCS}\label{Subs:Capua}

\begin{figure}[h]
\vspace{-0.2cm}
\begin{center}\includegraphics[trim = 0mm 8mm 10mm 14mm,clip,scale=0.6]{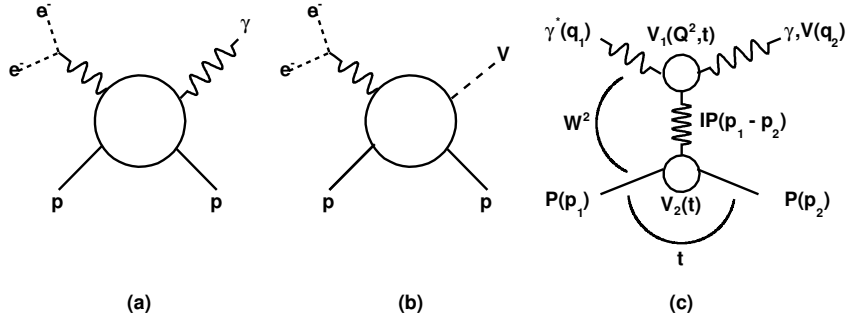}\end{center}
%  \vspace{-1.0cm}
\caption{ \label{fig:diagram} Diagrams of DVCS (\textbf{a}) and VMP (\textbf{b}); (\textbf{c}) DVCS (VMP) amplitude in a Regge-factorized form.}
\end{figure}

By Regge factorization,  Figure~\ref{fig:diagram}, the~DVCS amplitude can be written as
\begin{equation}\label{A1}
A(s,t,Q^2)_{\gamma^* p\rightarrow\gamma
p}=-A_0V_1(t,Q^2)V_2(t)(-is/s_0)^{\alpha(t)},
\end{equation}
where $A_0$ is a normalization factor, $V_1(t,Q^2)$ is the $\gamma^*
P \gamma$ vertex, $V_2(t)$ is the $p P p$ vertex and $\alpha(t)$ is the
exchanged pomeron trajectory, which we assume to be logarithmic:
\begin{equation}\label{alpha}
\alpha(t)=\alpha(0)-\alpha_1\ln(1-\alpha_2 t).
\end{equation}

Such a trajectory is nearly linear for small $|t|$, thus
reproducing the forward cone of the differential cross section, while its logarithmic asymptotic provides for the large-angle scaling behaviour, typical of hard collisions at
small distances, with~power-law fall-off in $|t|$, compatible with the quark counting rules. Here we are referring to the
dominant pomeron contribution eventually appended by a secondary trajectory, e.g.,~the $f$-Reggeon.

For convenience, and~following the arguments based on duality, the~$t$ dependence of the $p P p$ vertex is introduced via the $\alpha(t)$ trajectory:
$V_2(t)=e^{b \alpha(t)}$ where $b$ is a parameter.
A generalization of this concept will be
applied also to the upper, $\gamma^* P \gamma$ vertex by
introducing the  trajectory
\begin{equation}\label{beta}
\beta(z)=\alpha(0)-\alpha_1\ln(1-\alpha_2 z),
\end{equation}
where the value of the parameter $\alpha_2$ may be different in $\alpha(t)$ and $\beta(z)$ (a relevant check will be possible when more data are available).

Hence the scattering amplitude, with~correct signature, becomes
\begin{equation}\label{A2}
A(s,t,Q^2)_{\gamma^* p\rightarrow\gamma p}= -A_0e^{b\alpha(t)}e^{b
\beta(z)}(-is/s_0)^{\alpha(t)}= -A_0e^{(b+L)\alpha(t)+b\beta(z)},
\end{equation}
where $L\equiv\ln(-is/s_0)$.

The model contains a limited number of parameters. Moreover,
most of them can be estimated a priori. The~product
$\alpha_1\alpha_2$ is just the forward slope $\alpha'$ of the
Reggeon ($\approx$0.2~GeV$^{-2}$ for the pomeron, but~much higher
for $f$ and/or for an effective Reggeon).  The~value of $\alpha_1$ can be estimated from the wide-angle quark counting rules. For~large $t$ ($|t|>>$1~GeV$^2$)
the amplitude goes roughly as $\sim
e^{-\alpha_1\ln(-t)}=(-t)^{\alpha_1},$ where the power $\alpha_1$
is related to the number of quarks in a~collision.

From Equation~(\ref{A2}) the slope of the forward cone is
\begin{equation}\label{slope}
B(s,Q^2,t)=\frac{d}{dt}\ln|A|^2=2\left[b+\ln\left({s\over{s_0}}\right)\right]{\alpha'\over{1-\alpha_2
t}}+ 2b\,{\alpha'\over{1-\alpha_2 z}},
\end{equation}
which, in~the forward limit, $t=0$ reduces to
\begin{equation}\label{slope1}
B(s,Q^2)=2\left[b+\ln\left({s\over{s_0}}\right)\right]\alpha'+
2b\,{\alpha'\over{1+\alpha_2 Q^2}}.
\end{equation}

Thus, the~slope shows shrinkage in $s$ and antishrinkage in $Q^2.$

In the $Q^2\rightarrow 0$ limit the  Equation~(\ref{A2}) becomes
\begin{equation}\label{amplitude4}
A(s,t)=-A_0e^{2b\alpha(t)}(-is/s_0)^{\alpha(t)}
\end{equation}
where we recognize a typical Regge-behaved photoproduction
(or, for~$Q^2\rightarrow m_H^2,$ on-shell hadronic ($H$)) amplitude.
The related deep inelastic scattering structure function is
recovered by setting \mbox{$Q_2^2=Q^2_1=Q^2$ and $t=0$}, to~get a typical
elastic virtual forward Compton scattering amplitude:
\begin{equation}\label{amplitude5}
A(s,Q^2)=-A_0e^{b(\alpha(0)-\alpha_1\ln(1+\alpha_2
Q^2))}e^{(b+\ln(-is/s_0))\alpha(0)}\propto -(1+\alpha_2
Q^2)^{-\alpha_1}(-is/s_0)^{\alpha(0)}.
\end{equation}

In the Bjorken limit, when both $s$ and
$Q^2$ are large and $t=0$ (with $x\approx Q^2/s$ valid for large
$s$), the~structure function is given by:
\begin{equation}\label{F2}
F_2(s,Q^2)\approx {(1-x)Q^2\over{\pi\alpha_e}}\Im A(s,Q^2)/s,
\end{equation}
where $\alpha_e$ is the electromagnetic coupling constant and the
normalization is $\sigma_t(s)=\frac{4\pi}{s}\Im A(s,Q^2).$ The~resulting structure function has correct (required by gauge
invariance) $Q^2\rightarrow 0$ limit and approximate scaling (in
$x$) behavior for large enough $s$ and $Q^2$.

\section{Reggeometry}\label{Sec: Reggeomety}

In a series of papers (see Refs.~\cite{Salii1, Salii2}), photon virtuality was incorporated in a ``geometrical'' way, reflecting the observed trend in the decrease of the forward slope as a function of  $\widetilde {Q^2}$. This geometrical approach, combined with the Regge-pole model, was named ``Reg-ge-ometry'' (a wordplay, pun). A~Reggeometric amplitude dominated by a single pomeron  shows reasonable agreement with the HERA data on VMP and DVCS, when fitted separately to each~reaction.

As further step, to~reproduce the observed trend of hardening as $\widetilde{Q^2}$ increases, and~following Donnachie and Landshoff~\cite{DL1, DL2}, a~two-term amplitude, characterized by a two-component---``soft'' + ``hard''---pomeron,  was suggested~\cite{Salii1}. We stress that the pomeron is unique, but~we construct it as a sum of two terms. Then, the~amplitude is defined as
\begin{equation}
A(\widetilde {Q^2},s,t)=A_s (\widetilde {Q^2},s,t)+A_h(\widetilde {Q^2},s,t)
\label{two-term-amp}
\end{equation}
($s=W^2$ is the square of the c.m.s. %Please define if appropriate.
energy),
such that the relative weight of the two terms changes with $\widetilde {Q^2}$ in the right way, i.e.,~the ratio $r=A_h/A_s$ increases as the reaction becomes ``harder'' and v.v. It is interesting to note that this trend is not guaranteed ``automatically'': both the ``scaling'' model~\cite{Capua} or the Reggeometric one~\cite{ Salii1, Salii2} show the opposite tendency, that may not be merely an accident and whose reason should be better understood. This ``wrong'' trend can and should be corrected, and~in fact it was corrected~\cite{DL1, DL2} by means of additional $\widetilde {Q^2}$-dependent factors $H_i(\widetilde {Q^2}),\ i=s,h$ modifying the $\widetilde {Q^2}$
dependence of the amplitude,
%. These factors  have been chosen
in a such way as to provide increasing of the weight of the hard component with increasing $\widetilde {Q^2}$. To~avoid conflict with unitarity, the~rise with $\widetilde {Q^2}$ of the hard component is finite (or moderate), and~it terminates at some saturation scale,
%$\widetilde Q^2$,
whose value is determined phenomenologically. In~other words, the~``hard" component, invisible at small $\widetilde {Q^2}$, gradually takes over as $\widetilde {Q^2}$ increases. An~explicit example of these functions is presented~below.

%\subsubsection{Single-Component Reggeometric %Pomeron}
%\label{sec:Single}

Recall that the invariant scattering amplitude is defined as
\begin{equation}\label{Amplitude1}
A(Q^2,s,t)=\widetilde He^{-\frac{i\pi\alpha(t)}{2}}\left(\frac{s}{s_0}\right)^{\alpha(t)} e^{2\left(\frac{a}{\widetilde{Q^2}}+\frac{b}{2m_N^2}\right)t},
\end{equation}
where
\begin{equation}
\alpha(t)=\alpha_0+\alpha't
\end{equation}
is the linear pomeron trajectory (the use of more advanced, non-linear trajectories of Section~\ref{Subs:Trajectory} is straightforward), $a$ and $b$ are two parameters to be determined from the fitting procedure and $m_N$ is the nucleon mass. The~coefficient $\widetilde H$ is a function providing the right behavior of elastic cross sections in $\widetilde{Q^2}$:

$$
\widetilde H\equiv \widetilde H(\widetilde{Q^2})=\frac{\widetilde{A_0}}{\left(1+\frac{\widetilde{Q^2}}{{Q_0^2}}\right)^{n_s}},
$$
where $\widetilde{A_0}$ is a normalization factor, $Q_0^2$ is a scale for the virtuality and $n_s$ is a real positive~number.

In this model one uses an effective pomeron, which can be ``soft'' or ``hard'', depending on the reaction and/or kinematic region defining its ``hardness''. In~other words, the~ values of the parameters $\alpha_0$ and $\alpha'$ must be fitted to each set of the data. Apart from $\alpha_0$ and $\alpha'$, the~model contains five more sets of free parameters, different in each reaction, as~shown in Table~\ref{tab:one_term}.
The exponent in the exponential factor in Equation~(\ref{Amplitude1}) reflects the geometrical nature of the model: $a/\widetilde {Q^2}$  and $b/2m_N^2$ correspond to the ``sizes" of upper and lower vertices in Figure~\ref{fig:diagram}c.

With the norm
%mdpi: There is no find {eq:norm} in this paper, %please add.
\begin{equation}\label{eq:dcsdt_from_Amplitude}
\frac{d\sigma_{el}}{dt}=\frac{\pi}{s^2}|A(Q^2,s,t)|^2,
\end{equation}
the differential and integrated elastic cross sections become,
\begin{equation}\label{eq:dcsdt_1}
\frac{d\sigma_{el}}{dt}=\frac{A_0^2}{\left(1+\frac{\widetilde{Q^2}}{{Q_0^2}}\right)^{2n}}\left(\frac{s}{s_0}\right)^{2(\alpha(t)-1)}e^{4\left(\frac{a}{\widetilde{Q^2}}+\frac{b}{2m_N^2}\right)t}
\end{equation}
and
\begin{equation}\label{eq:cs_1}
\sigma_{el}=\frac{A_0^2} {\left(1+\frac{\widetilde{Q^2}}{{Q_0^2}}\right)^{2n}}
\frac{\left(\frac{s}{s_0}\right)^{2(\alpha_0-1)}}
{4\left(\frac{a}{\widetilde{Q^2}}+\frac{b}{2m_N^2}\right)+2\alpha'\ln\left(\frac{s}{s_0}\right)},
\end{equation}
where
$$A_0=-\frac{\sqrt{\pi}}{s_0}\widetilde{A_0}.$$

\begin{figure}[h]
\vspace{-0.2cm}
\centering
\includegraphics[trim = 0mm 0mm 1mm 0mm,clip,scale=0.34]{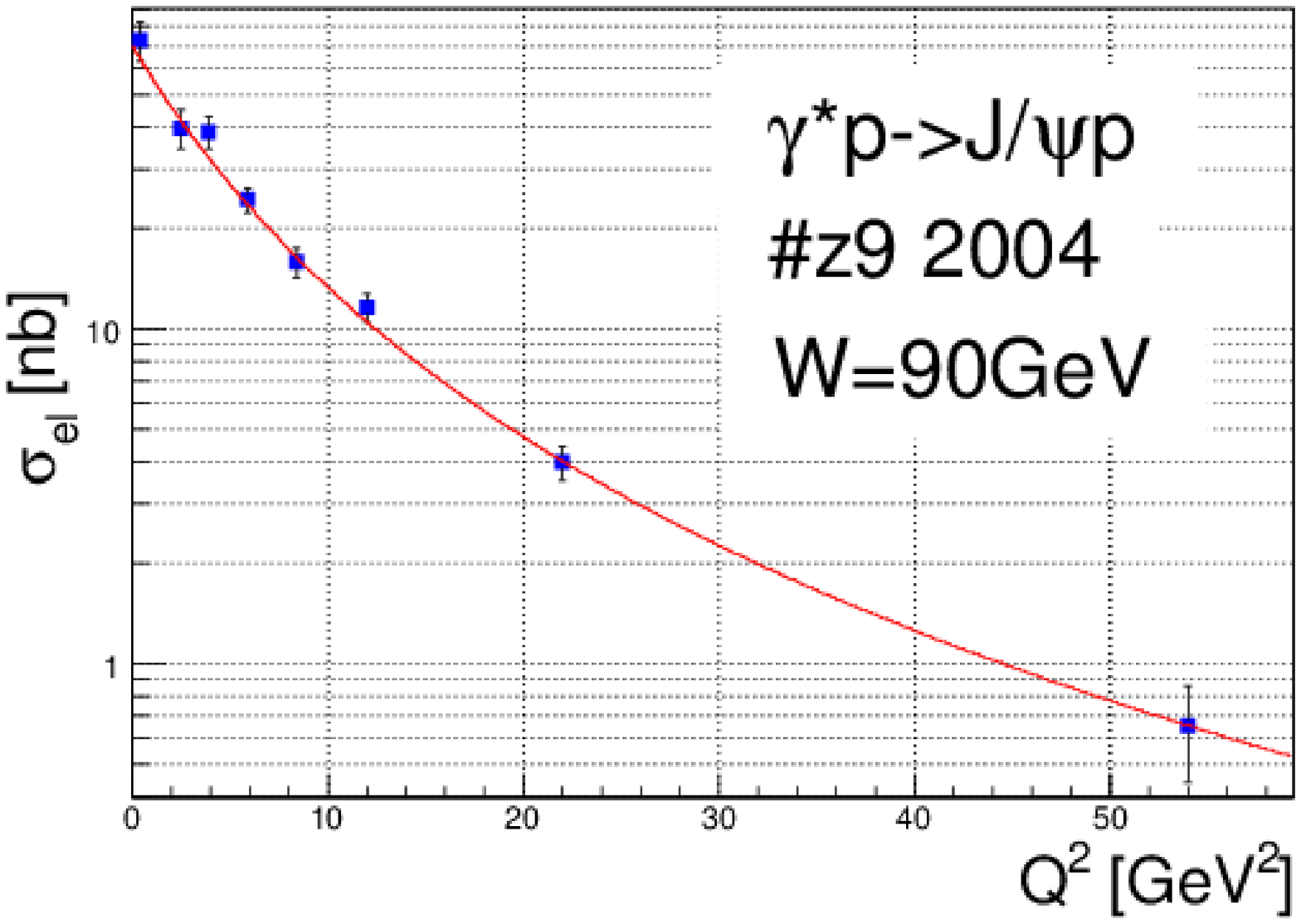}
\includegraphics[trim = 1mm 0mm 1mm 0mm,clip,scale=0.72]{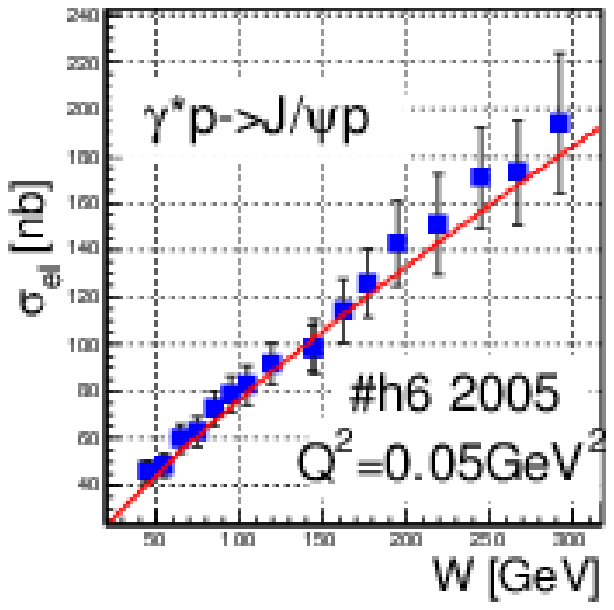}
% \includegraphics[trim = 1mm 2mm 1mm 0mm,clip,scale=0.72]{dcsdt.eps}
%   \includegraphics[trim = 3mm 0mm 13mm 6mm,clip,scale=0.34]{cs_Q2.eps}
%   \includegraphics[trim = 4mm 74mm 135mm 0mm,clip,scale=0.75]{cs_W.eps}
%   \includegraphics[trim = 69mm 72mm 67mm 0mm,clip,scale=0.75]{dcsdt.eps}
%  \vspace{-1.0cm}
\caption{ \label{fig:onetermJpsi}Representative fits of Equations~\eqref{eq:dcsdt_1} and \eqref{eq:cs_1} to the data on $J/\psi$ production. The~values of the fitted parameters are compiled in Table~\ref{tab:one_term}.}
\end{figure}

The results of the fits are presented in Table~\ref{tab:one_term}. The~``mass parameter'' for DVCS was set to $M=0$~GeV, therefore in this case $\widetilde{Q^2}=Q^2$. Each type of reaction was fitted separately.
As it can be seen from the right plot of Figure~\ref{fig:onetermJpsi}, the~single-term model fails to fit both the high- and low-$|t|$ regions properly, especially when soft (photoproduction or low $Q^2$) and hard (electroproduction or high $Q^2$) regions are considered.
%As it can be seen from the right diagram of Figures~\ref{fig:oneterm_rho},
One of the problems of the single-term Reggeometric pomeron model, Equation~\eqref{Amplitude1}, is that the fitted parameters in this model acquire particular values for each reaction, %which is not economic, and~which is one of the reasons for its extension to two terms  (next Section). %At the same time, they are useful and instructive in performing the more complicated fit of the universal model, to~be presented in the next Section.
%
%### Figure~2 ######
\unskip

%\end{document}

% Table~1
\begin{table}[h]
\centering
\footnotesize
\caption{Values of the parameters in Equations~(\ref{eq:dcsdt_1}) and (\ref{eq:cs_1}) fitted to data on vector meson production (VMP) and DVCS at HERA. %Please define if appropriate.
The parameters with indefinite error bars were fixed at the fitting stage.
\label{tab:one_term}}
\scalebox{.9}[0.9]{
\begin{tabular}{ccccccccc}\toprule
&\boldmath{$A_0$} $\left[\frac{\sqrt\text{\textbf{nb}}}{\text{\textbf{GeV}}}\right]$
&\boldmath{$\widetilde{Q^2_0}$} \boldmath{$\left[\text{\textbf{GeV}}^\textbf{2}\right]$}&  \boldmath{ $n$}
&\boldmath{$\alpha_{0}$}& \boldmath{$\alpha'$}  \boldmath{$\left[\frac{1}{\text{GeV}^{\textbf{2}}}\right]$}
&\boldmath{$a$}&\boldmath{$b$}&\boldmath{${\tilde\chi}^2$} \\ \midrule
%           $pp$ &5.9$\pm$3.3&$***$              &  0.00        &1.05$\pm$0.08& 0.28$\pm$0.28& 2.9 $\pm$1.6 &0.00        &1.53\\ %  ine
$\rho^0$&344 $\pm$ 376&0.29 $\pm$ 0.14      &1.24 $\pm$ 0.07 &1.16 $\pm$ 0.14& 0.21 $\pm$ 0.53& 0.60 $\pm$ 0.33&0.9 $\pm$ 4.3 &2.74\\ %  ine
$\phi$  &58 $\pm$ 112 &0.89 $\pm$ 1.40      &1.30 $\pm$ 0.28 &1.14 $\pm$ 0.19& 0.17 $\pm$ 0.78& 0.0 $\pm$ 19.8 &1.34 $\pm$ 5.09&1.22\\%  ine
$J/\psi$& 30 $\pm$ 31 &2.3 $\pm$ 2.2        &1.45 $\pm$ 0.32 &1.21 $\pm$ 0.09& 0.077 $\pm$ 0.072& 1.72       &1.16        &0.27\\  %  ine
$\varUpsilon$(1S)&37 $\pm$ 100&0.93 $\pm$ 1.75      &1.45 $\pm$ 0.53 &1.29 $\pm$ 0.25& 0.006 $\pm$ 0.6& 1.90         &1.03        &0.4\\  %  ine
%        $\gamma$ &8.8$\pm$8.7&0.54$\pm$0.66      &0.93$\pm$0.17 &1.23$\pm$0.09& 0.05$\pm$0.08& 1.6          &1.9         &1.05\\
$DVCS$&14.5 $\pm$ 41.3&0.28 $\pm$ 0.98     &0.90 $\pm$ 0.18 &1.23 $\pm$ 0.14& 0.04 $\pm$ 0.71& 1.6          &1.9 $\pm$ 2.5 &1.05\\ \bottomrule
\end{tabular}}
\end{table}

%\end{document}

% {\bf To describe $pp$ scattering one soft Pomeron is enough%
% \footnote{in the recent article [arXiv 1112:2485] Donnachie (???) and Landshoff state that elastic $pp$ scattering contains (exhibit) also small fraction of hard Pomeron exchange(?).}.
% }

The VMP results clearly show hardening of the pomeron in the change of $\alpha_0$ and $\alpha'$ when going from light to heavy vector~mesons.

It is also interesting to note that the effective pomeron trajectory for DVCS ($\alpha_0 = 1.23,$ $\alpha'=0.04$, see Table~\ref{tab:one_term}), contrary to expectations is rather ``hard''.
%### Figure~4 ######
%\begin{figure}[!ht]
%\vspace{-0.2cm}
%\centering
%\includegraphics[trim = 0mm 0mm 0mm 0mm,clip,scale=0.40]{exp.eps}
%  \vspace{-1.0cm}
%\caption{ \label{fig:two_exp} In the case of a two-component Pomeron, the $t$-dependence of the scattering amplitude will have a form $A(t)=A_1e^{B_1t}+A_2e^{B_2t}$, thus in the region where both ``soft'' and ``hard'' terms are of the same order (transition region), a break in the plot of the differential cross section appears. This situation is typical for the $\rho^{0}$ data, especially in the photoproduction region (see Figure~\ref{fig:oneterm_rho}). In this plot show only the sum of two exponents, but in the case of the differential cross sections $\frac{d\sigma_el}{dt}=\frac{\pi}{s^2}|A_s+A_f|^2$ the interference term also should be added.}
%\end{figure}
%---------------------------------------------------------------------

\subsection*{Two-Component Reggeometric~Pomeron}
\label{sec:Two-components model}
%\subsection{Amplitude with two, ``soft'' and %``hard'', components}
Now we introduce the universal, ``soft'' and ``hard'', pomeron model.
Using the Reggeometric ansatz of Equation~\eqref{Amplitude1}, we write the amplitude as a sum of two parts, corresponding to the ``soft'' and ``hard'' components of a universal, unique  pomeron:
\begin{equation}\label{eq:Amplitude_hs_prime}
A(Q^2,s,t)=
\widetilde{H_s}\,e^{-i\frac{\pi}{2}\alpha_s(t)}\left(\frac{s}{s_{0s}}\right)^{\alpha_s(t)} e^{2\left(\frac{a_{s}}{\widetilde{Q^2}}+\frac{b_{s}}{2m_N^2}\right)t}
+\widetilde{H_h}\,e^{-i\frac{\pi}{2}\alpha_h(t)}\left(\frac{s}{s_{0h}}\right)^{\alpha_h(t)} e^{2\left(\frac{a_{h}}{\widetilde{Q^2}}+\frac{b_{h}}{2m_N^2}\right)t}.
\end{equation}

Here  $s_{0s}$ and $s_{0h}$
are squared energy scales, and~$a_i$ and $b_i$, with~$i = s,h$, are parameters to be determined with the fitting procedure. The~two coefficients $\widetilde{H_s}$ and $\widetilde{H_h}$ are functions similar to those defined in Refs.~\cite{DL1, DL2}:
\begin{equation}\label{eq:HsHh_tilde}
\widetilde{H_s}\equiv\widetilde{H_s}(\widetilde{Q^2})=\frac{\widetilde{A_s}}{{\Bigl(1+\frac{\widetilde{Q^2}}{{Q_s^2}}\Bigr)}^{n_s}}, ~~~~~~\quad
\widetilde{H_h}\equiv\widetilde{H_h}(\widetilde{Q^2})=\frac{\widetilde{A_h} \Bigl(\frac{\widetilde{Q^2}}{Q_h^2}\Bigr)}{{\Bigl(1+\frac{\widetilde{Q^2}}{{Q_h^2}}\Bigr)}^{n_h+1}},
%\frac{ \widetilde{A_h} \Bigl(\frac{ \widetilde{Q^2}{\{Q_h^2}}\Bigr)}{{\Bigl(1+\frac{\widetilde{Q^2}}{{Q_h^2}}\Bigr)}^{n_h+1}}.
\end{equation}
where $\widetilde{A_s}$ and $\widetilde{A_h}$ are normalization factors, $Q_s^2$ and $Q_h^2$ are scales for the virtuality, $n_s$ and $n_h$ are real positive numbers.
Each component of Equation~\eqref{eq:Amplitude_hs_prime} has its own, ``soft'' or ``hard'', Regge (here pomeron) trajectory:
$$\alpha_s(t)=\alpha_{0s}+\alpha_s't, ~~~~~~~~~~~\quad  \alpha_h(t)=\alpha_{0h}+\alpha_h't.$$

The terms ``soft'' and ``hard'' may be misleading, alluding to the
wide-spead nomenclature by which the ``soft'' pomeron is associated with the traditional Regge-pole theory, while the ``hard'' one is the BFKL pomeron derived from QCD, having little to do with the former, by~meaning that they are different object. Instead, we insist that there is only one pomeron, although~it may be complicated, consisting of more terms etc. In~our case the two terms refer to a single object, THE pomeron, while its (two) parts are weighted with virtuality $Q^2$.

As an input we use the parameters suggested by Donnachie and Landshoff~\cite{DL1, DL2}, so that
$$\alpha_s(t) = 1.08 + 0.25t,~~~~~~~~~~\quad \alpha_h(t) = 1.40 + 0.1t.$$
%The factor $\frac{s_0}{\sqrt{\pi}}$ in front of the amplitude is convenient just for normalization reasons.

The ``pomeron''  amplitude (\ref{eq:Amplitude_hs_prime}) is unique, valid for all diffractive
reactions, its ``softness'' or ``hardness'' depending on the relative $\widetilde{Q^2}$-dependent weight of the two components, governed by the relevant factors $\widetilde H_s(
\widetilde Q^2)$ and $\widetilde {H_h}(\widetilde {Q^2)}$.
%here $\widetilde{H_{\;}}_{\!\!s,h}\equiv\widetilde{H_{\;}}_{\!\!s,h}(\widetilde{Q^2})$
%In our earlier papers~\cite{Acta}  (see also in Equation~(\ref{Amplitude1})) the parameter $b_i,\ \ i=s,h$ was of "geometric" form
%$$2\left(\frac{a_{s,h}}{\widetilde{Q^2}}+\frac{b_{s,h}}{2m_p^2}\right), \eqno{(*)}$$
%where the first term specified the "hardness" of the reaction (upper vertex in Figure~\ref{fig:diagram}) and the second, constant term corresponds to the lower vertex of that diagram (for details see Refs.~\cite{Acta}).

To reduce the number of free parameters, we simplified the model, by~fixing $a_{s,h}=0$ and substituting the exponent $2\left(\frac{a_{s,h}}{\widetilde{Q^2}}+\frac{b_{s,h}}{2m_N^2}\right)$ with $b_{s,h}$ in Equation~\eqref{eq:Amplitude_hs_prime}. The~proper variation with $\widetilde {Q^2}$ will be provided by the factors $\widetilde{H_s}(\widetilde{Q^2})$ and $\widetilde{H_h}(\widetilde{Q^2})$.

%%%%%%%%%%%%%======================
Consequently, the~scattering amplitude assumes the form
\begin{equation}\label{eq:Amplitude_hs}
A(s,t,Q^2,M_V^2)=
\widetilde{H_s}\,e^{-i\frac{\pi}{2}\alpha_s(t)}\left(\frac{s}{s_{0s}}\right)^{\alpha_s(t)} e^{b_st}
+\widetilde{H_h}\,e^{-i\frac{\pi}{2}\alpha_h(t)}\left(\frac{s}{s_{0h}}\right)^{\alpha_h(t)} e^{b_ht}.
\end{equation}

The ``Reggeometric'' combination $2\left(\frac{a_{s,h}}{\widetilde{Q^2}}+\frac{b_{s,h}}{2m_N^2}\right)$ was important for the description of the slope $B(Q^2)$ within the single-term pomeron model (see previous Section), but~in the case of two terms the $Q^2$-dependence of $B$ can be reproduced without this extra combination, since each term in the amplitude \eqref{eq:Amplitude_hs} has its own $Q^2$-dependent factor $\widetilde{H_{\ }}_{\!\!s,h}(Q^2)$.

By using the amplitude (\ref{eq:Amplitude_hs}) and Equation~\eqref{eq:dcsdt_from_Amplitude}, we calculate the differential and elastic cross sections, by~setting for simplicity $s_{0s} = s_{0h} = s_0$, to~obtain
\begin{equation}\label{eq:dcsdt(h+s)}
\frac{d\sigma_{el}}{{dt}}=H_s^2e^{2\{L(\alpha_s(t)-1)+{b_s}t\}}+H_h^2e^{2\{L(\alpha_h(t)-1)+{b_h}t\}}
\end{equation}

$$+2H_sH_he^{\{L(\alpha_s(t)-1)+L(\alpha_h(t)-1)+({b_s}+{b_h})t\}}\cos\Bigl(\frac{\pi}{2}(\alpha_s(t)-\alpha_h(t))\Bigr),$$
\begin{equation}\label{eq:cs(h+s)}
\sigma_{el}=\frac{H_s^2e^{2\{L(\alpha_{0s}-1)\}}}{2(\alpha_s'L+{b_s})}
+\frac{H_h^2e^{2\{L(\alpha_{0h}-1)\}}}{2(\alpha_h'L+{b_h})}
+2H_sH_he^{L(\alpha_{0s}-1)+L(\alpha_{0h}-1)}
\frac{\mathfrak{B}\cos\phi_0+\mathfrak{L}\sin\phi_0} {\mathfrak{B}^2 + \mathfrak{L}^2}.
\end{equation}

In these two equations we use the notations
\begin{equation}\label{eq:Denotation}
\begin{array}{l}
L=\ln\left({s/s_{0}}\right),
\\\phi_0=\frac{\pi}{2}(\alpha_{0s}-\alpha_{0h}),
\end{array}\qquad
\begin{array}{l}
\mathfrak{B}=L\alpha_s' + L\alpha_h'+({b_s}+{b_h}),
\\\mathfrak{L}=\frac{\pi}{2}(\alpha_s'-\alpha_h'),
\end{array}
\nonumber
\end{equation}
\begin{equation}%\label{eq:HsHh}
H_s(\widetilde{Q^2})=\frac{A_s}{{\Bigl(1+\frac{\widetilde{Q^2}}{{Q_s^2}}\Bigr)}^{n_s}}, \quad
H_h(\widetilde{Q^2})=\frac{A_h\Bigl(\frac{\widetilde{Q^2}}{{Q_h^2}}\Bigr)}{{\Bigl(1+\frac{\widetilde{Q^2}}{{Q_h^2}}\Bigr)}^{n_h+1}}, \nonumber
\end{equation}
with
$$A_{s,h}=-\frac{\sqrt{\pi}}{s_{0}}\widetilde{A_{\ }}_{\!\!s,h}.$$

%%%####  \lnH-form #####
Notice that amplitude~(\ref{eq:Amplitude_hs}) can be rewritten in the form
$$
A(s,t,Q^2,{M_v}^2)= \widetilde{A_s}e^{-i\frac{\pi}{2}\alpha_s(t)}\left(\frac{s}{s_{0}}\right)^{\alpha_s(t)}
e^{b_st - n_s\ln{\left(1+\frac{\widetilde{Q^2}}{\widetilde{Q_s^2}}\right)}}
$$
\begin{equation}
+\widetilde{A_h}e^{-i\frac{\pi}{2}\alpha_h(t)}\left(\frac{s}{s_{0}}\right)^{\alpha_h(t)}
e^{b_ht - (n_h+1)\ln{\left(1+\frac{\widetilde{Q^2}}{\widetilde{Q_h^2}}\right)}
+\ln{\left(\frac{\widetilde{Q^2}}{\widetilde{Q_h^2}}\right)} },
\label{eq:Amplitude_hs_modif}
\end{equation}
where  the two exponential factors $e^{b_st - n_s\ln{\left(1+\frac{\widetilde{Q^2}}{\widetilde{Q_s^2}}\right)}}$ and $e^{b_ht - (n_h+1)\ln{\left(1+\frac{\widetilde{Q^2}}{\widetilde{Q_h^2}}\right)}
+\ln{\left(\frac{\widetilde{Q^2}}{\widetilde{Q_h^2}}\right)}}$ can be interpreted as the product of the form factors of upper and lower vertices (see Figure~\ref{fig:diagram}c). Interestingly,~the~amplitude~\eqref{eq:Amplitude_hs_modif} resembles the scattering amplitude of Ref.~\cite{Capua}.
%%%########################

%------------------------------------------------------------------
%\newline
\subsubsection*{Fitting the Two-Component Pomeron to VMP and DVCS HERA~Data}

The fitting strategy is based on the minimization of the quantity $\tilde\chi^2=\frac{1}{N}\sum_{i=1}^{N} {\tilde\chi_i^2}$,
%where $\tilde\chi_i$ -- correspond to one class of reaction, and
where $N$ is the number of all reactions involved (i.e., $\rho$, $\phi$, $\omega$, $J/\psi$, $\varUpsilon$ and $\gamma$ production);
$\tilde\chi_i$ is the mean value of $\chi^2$ for different types of data for  selected class of reactions, defined as $\tilde\chi_i=\frac{1}{N_i}\sum_{k=1}^{N_i} {\tilde\chi_{k,i}^2}$,
where $\tilde\chi_{k,i}^2$  is $\chi_{k,i}^2/d.o.f.$ for i-th class of reactions and k-th type of data, i.e.,~those relative to $\sigma_{el}(Q^2)$, $\sigma_{el}(W)$ and ${d\sigma_{el}(t)}/{dt}$;
$N_i$ is number of different type of data for i-th class of~reactions.

The normalization parameters were fixed at
\begin{equation}\label{eq:f_i}
f_{\rho}=0.680,\ f_{\phi}=0.155,\ f_{\omega}=0.068,\ f_{J/\psi}=1,\ f_{\varUpsilon}=0.750
\end{equation}
and $s_{0}$ was set $1$\;GeV$^2$.

%\paragraph{DVCS}

DVCS and VMP are similar in the sense that in both reactions a vector particle is produced. However there are differences between the two because of the vanishing rest mass of the produced real photon. The~unified description of these two types of related reactions does not work by simply setting $M_{\gamma}=0.$  From fits we found $M_{DVCS}^{eff}=1.8$~GeV and a normalization factor $f_{DVCS}=0.091$ follows.

% For DVCS we introduce an ``effective mass'' $M_{DVCS}^{eff}=1.8$~GeV {\it for \bf (of) the real photon}.
% {\bf It is purely a phenomenological result from the fit. To describe the DVCS reactions together with the VMP reactions we need to introduce this non-zero ``mass'' for DVCS.
% Generally DVCS, maybe, should be treated separately from VMP, because our fits do not seem to be excellent for DVCS.
% We know that in VMP reactions the virtual foton split into quark anti-quark pair ($q\bar{q}$), which interact with the proton. In DVCS the same process take place, moreover it can split not only into one particular pair of quarks but into any of them. This may be an origin for this ``effective mass''.
% The normalization factor of DVCS was set to $f_{DVCS}=0.091$, it also was found form the fit.}
% %Both numbers were found from the fit to the data). Such a rescale enables us to describe DVCS together with VMP.

\begin{figure}[h] \centering
\includegraphics[trim = 0mm 3mm 5mm 0mm,clip, scale=0.71]{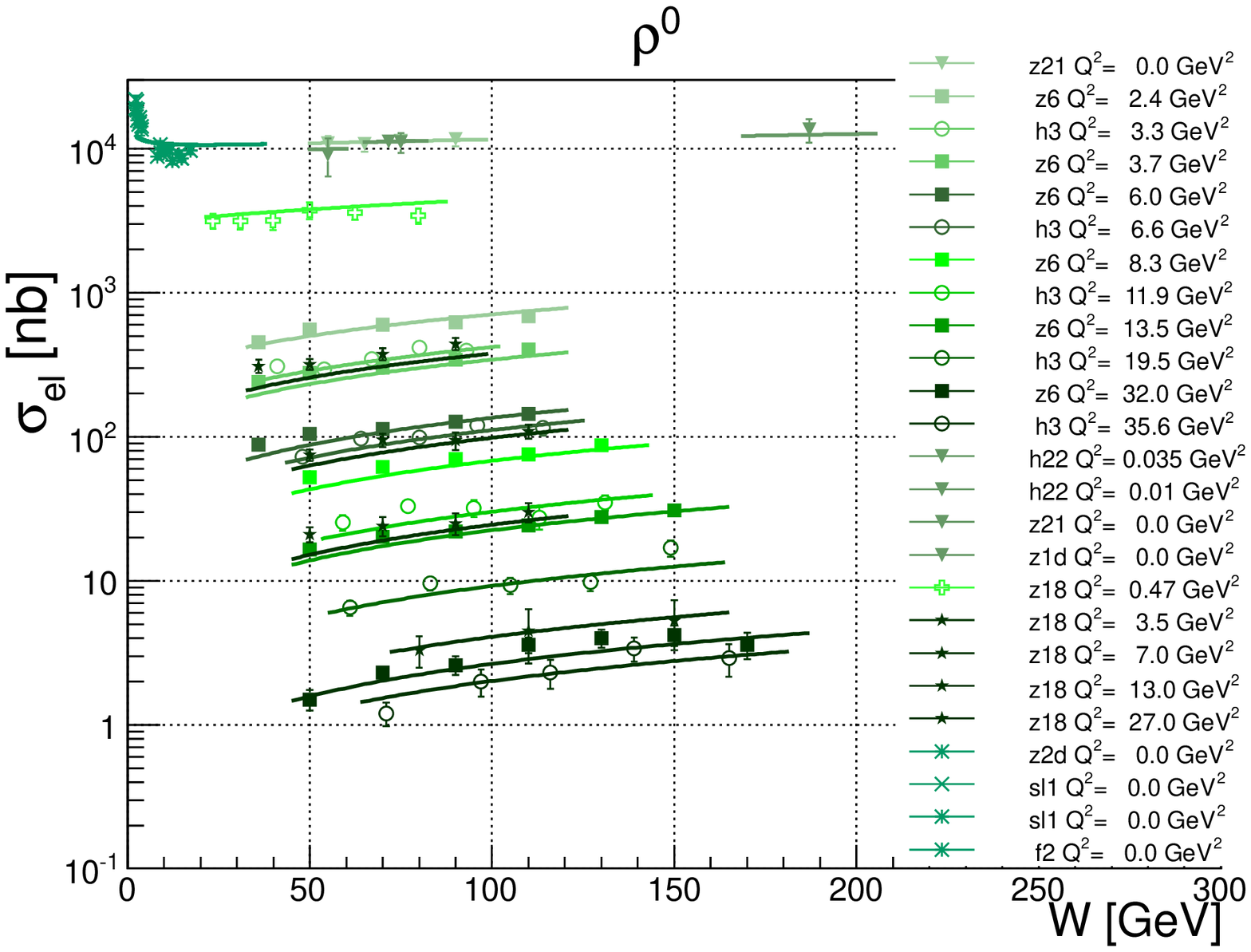} %\vspace{-0.4cm}
\caption{ \label{fig:cs_rho1(W)} \textls[-25]{Fit of Equation~(\ref{eq:cs(h+s)}) to  data on elastic cross section $\sigma_{el}(W)$ for $\rho^0$, for~different values of $Q^2$}.}
\end{figure}

\begin{figure}[h] \centering
\includegraphics[trim = 0mm 3mm 10mm 2mm,clip, scale=0.71]{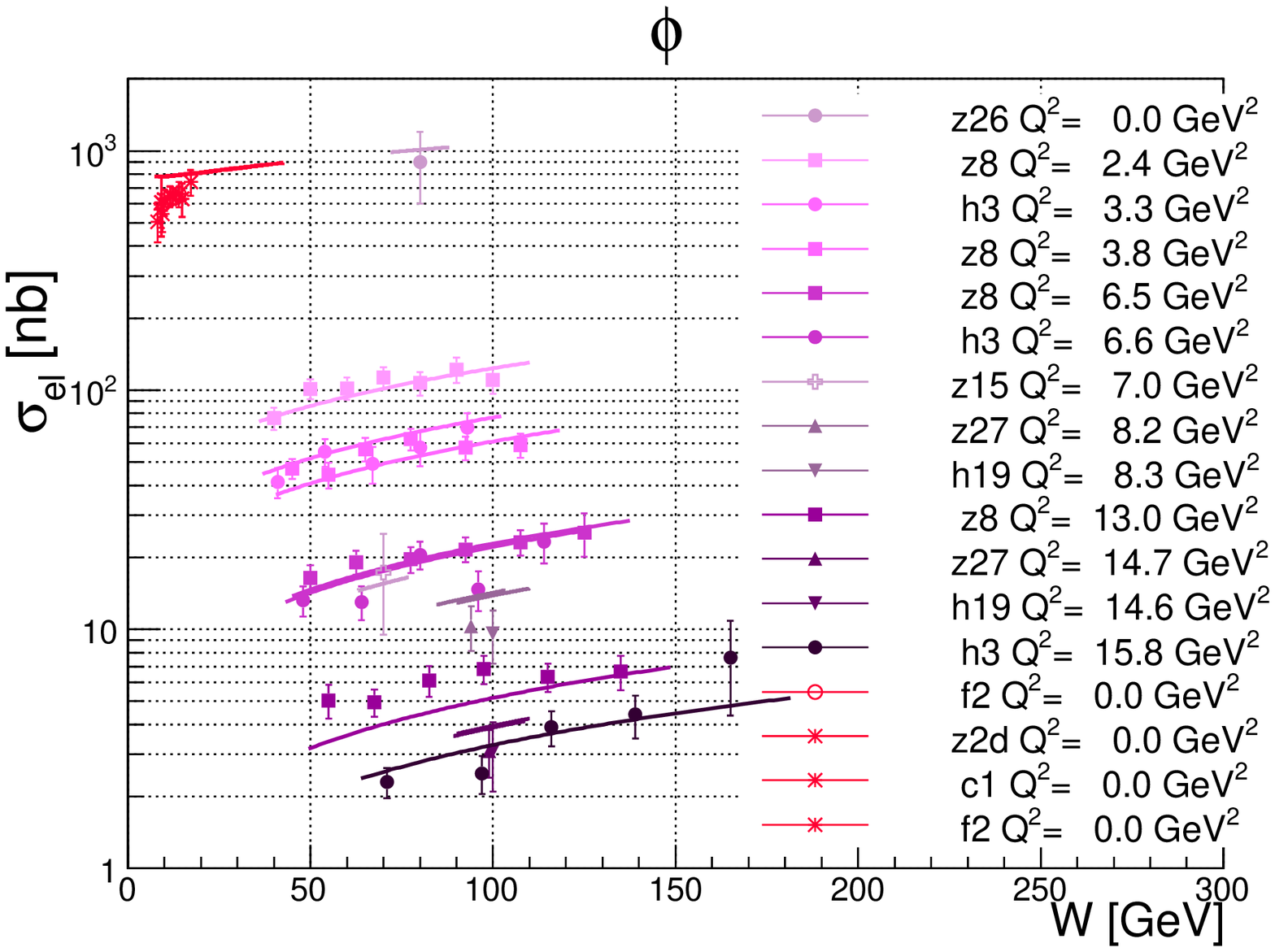}
\caption{ \label{fig:cs_phi(W)} \textls[-25]{Fit of Equation~(\ref{eq:cs(h+s)}) to the data on elastic cross section $\sigma_{el}(W)$  for $\phi$, for~different values of $Q^2$}.}
\end{figure}

\begin{figure}[h]  \centering
\includegraphics[trim = 0mm 3mm 10mm 2mm,clip, scale=0.71]{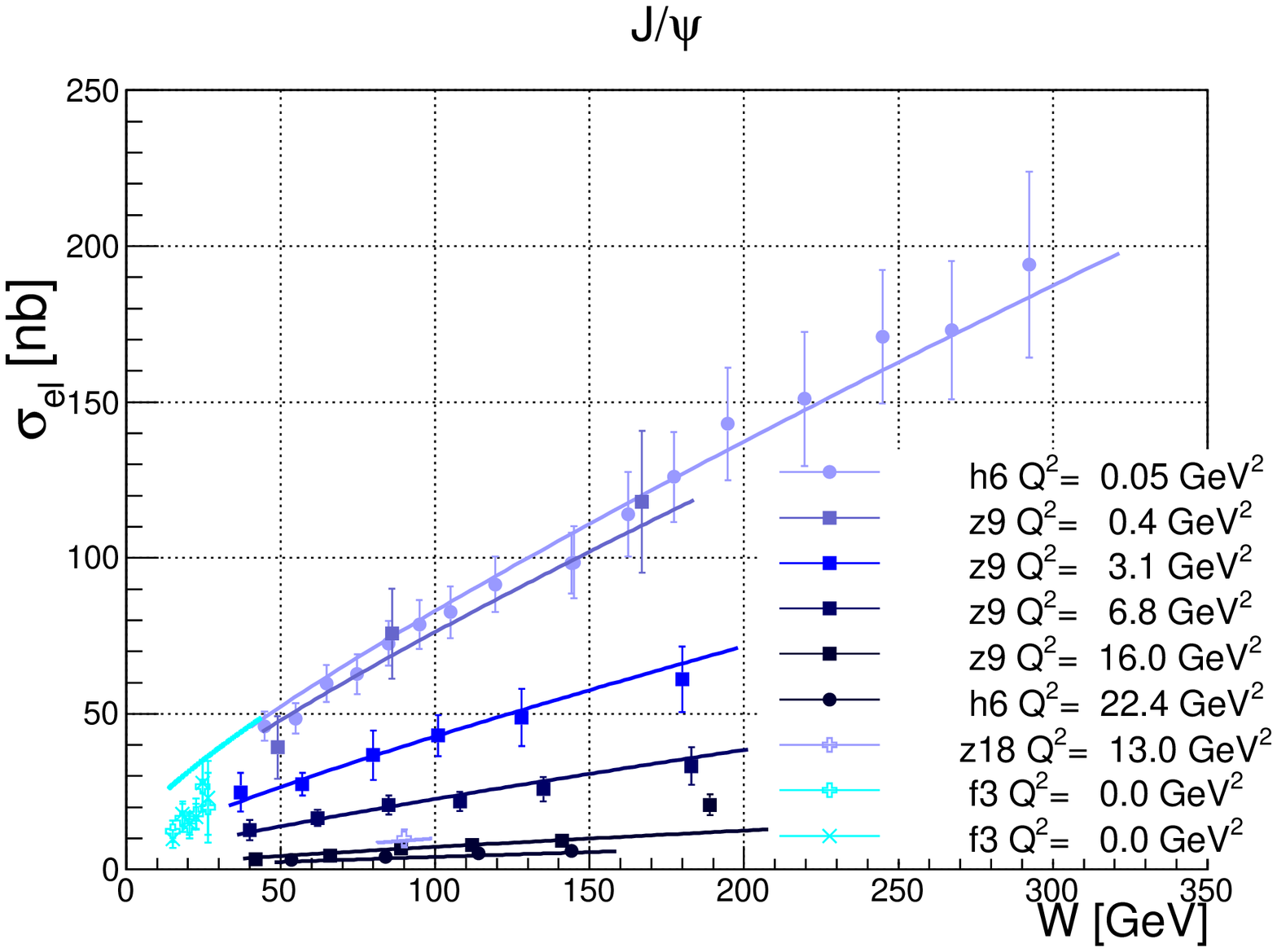}
\caption{ \label{fig:cs_Jpsi(W)} Fit of Equation~(\ref{eq:cs(h+s)}) to the data on elastic cross section $\sigma_{el}(W)$  for $J/\psi$, for~different values of $Q^2$.}
\end{figure}

\begin{figure}[h]  \centering
\includegraphics[trim = 0mm 3mm 12mm 2mm,clip, scale=0.75]{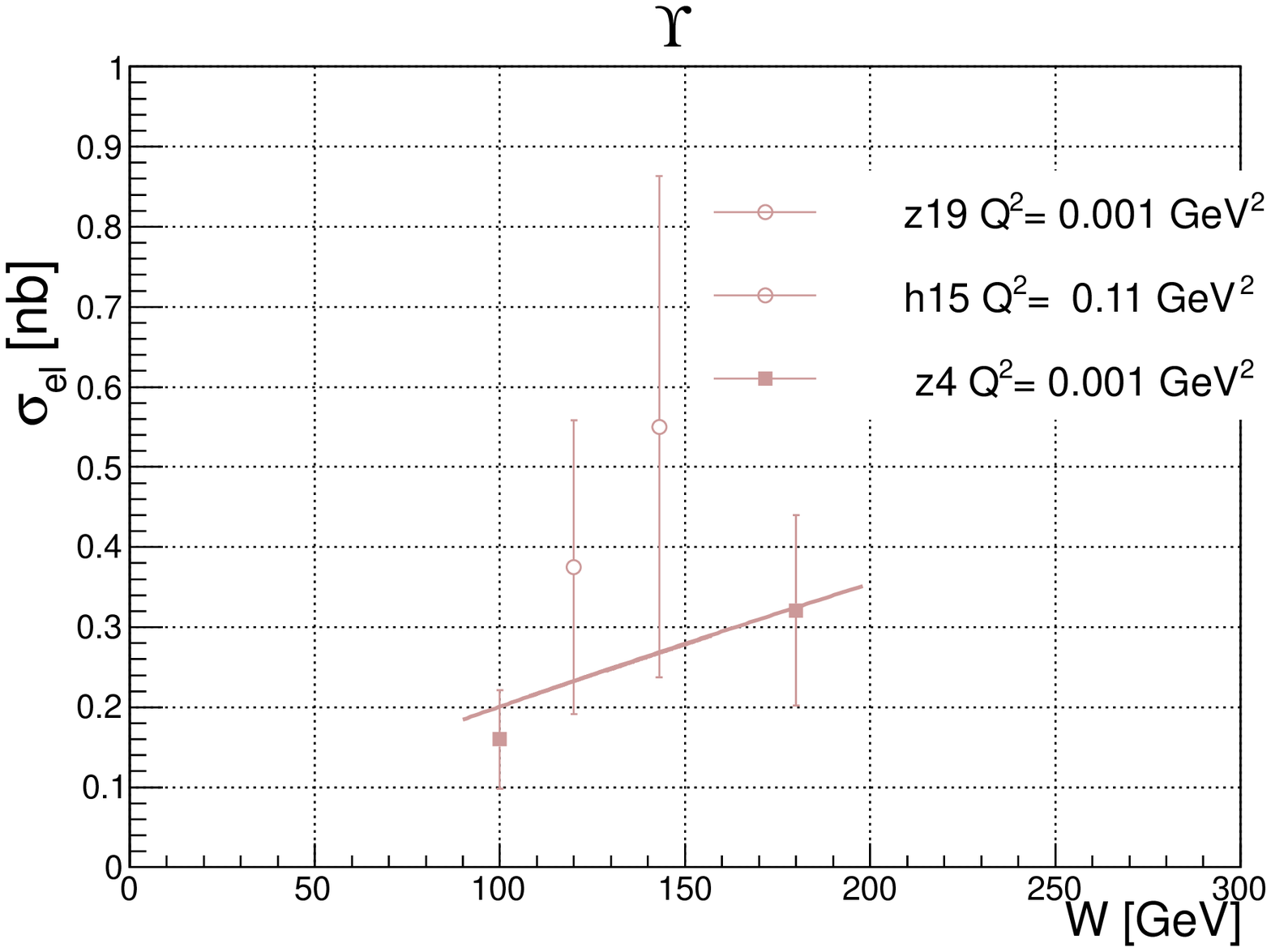}
\caption{ \label{fig:cs_Ups(W)} Fit of Equation~(\ref{eq:cs(h+s)}) to the data on elastic cross section $\sigma_{el}(W)$  for $\varUpsilon$, for~different values of $Q^2$.}
\end{figure}

\begin{figure}[h]\centering
\includegraphics[trim = 0mm 0mm 0mm 0mm,clip, scale=0.75]{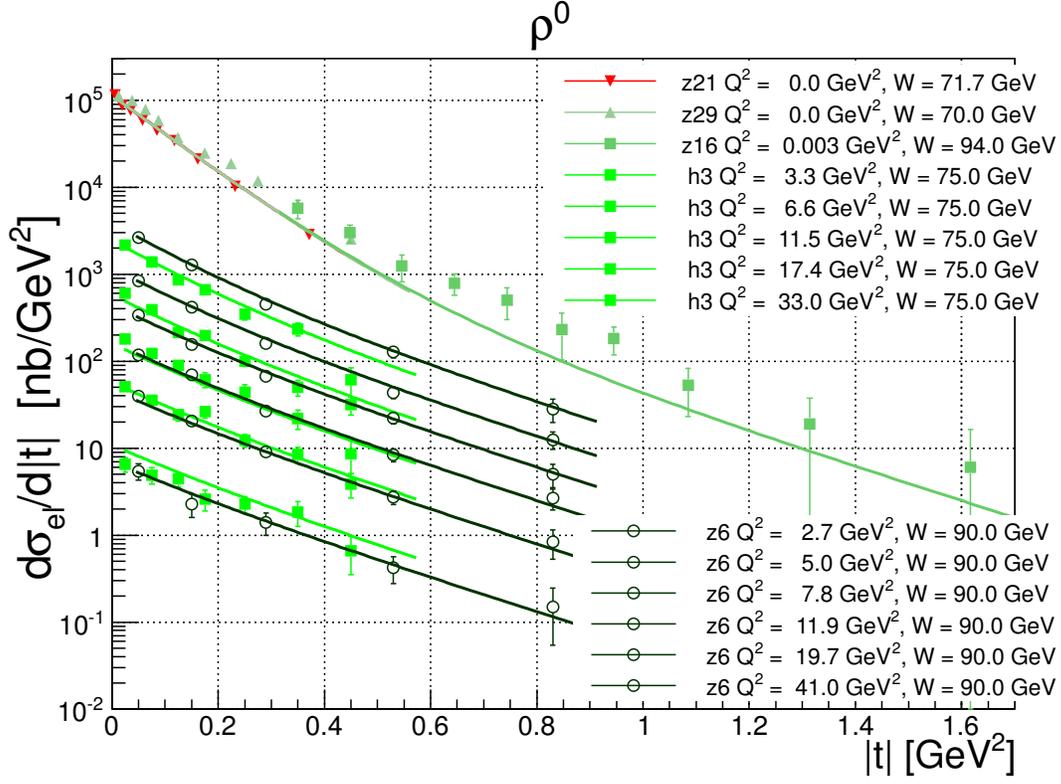}
\caption{ \label{fig:dcsdt_rho} Fit of Equation~(\ref{eq:dcsdt(h+s)}) to the data on the differential elastic cross section $d\sigma_{el}/dt$ for $\rho^0$, for~different values of $Q^2$ and $W$.}
\end{figure}

Some results of the fits are shown in Figures~\ref{fig:cs_rho1(W)}--\ref{fig:cs_Ups(W)} (for $\sigma_{el}(W)$) and Figure~\ref{fig:dcsdt_rho} (for $d\sigma_{el}(t)/dt$) for vector meson production,  with~the values of the fitted parameters given in Table~\ref{tab:fit1(s+h)}.
The mean value of the total $\tilde\chi^2$ (see above its definition) is equal to 0.986. The~mean values of $\tilde\chi^2$ of the fit for different observables (i.e., $\sigma_{el}(Q^2)$, $\sigma_{el}(W)$ or $d\sigma_{el}(t)/{dt})$ and different reactions (VMP or DVCS), together with the numbers of degrees of freedom (number of data points) and the global mean value $\tilde\chi_i^2$, are shown in Table~\ref{tab:fit1_chi}.

In Table~\ref{tab:fixed_trajec} the  parameters of the two-component pomeron model (Equations~(\ref{eq:dcsdt(h+s)}) and (\ref{eq:cs(h+s)})) fitted to the combined VMP and DVCS data are quoted, when the pomeron trajectories are fixed to \mbox{$\alpha_{s}(t)=1.08+0.25t$} and $\alpha_{h}(t)=1.20+0.01t$.

%and parametrs of Pomeron trajectories, we used, were:
%\begin{eqnarray}
%&\alpha_{0s}=1.091,&\quad \alpha'_{s}=0.46 \text{\qquad (sort Pomeron)};\\
%&\alpha_{0h}=1.214\pm0.092,&\quad \alpha'_{h}=0.10 \text{\qquad (hard Pomeron)}. \nonumber
%\end{eqnarray}

\begin{figure}[h]
\centering
\includegraphics[trim = 0mm 0mm 10mm 3mm,clip, scale=0.75]{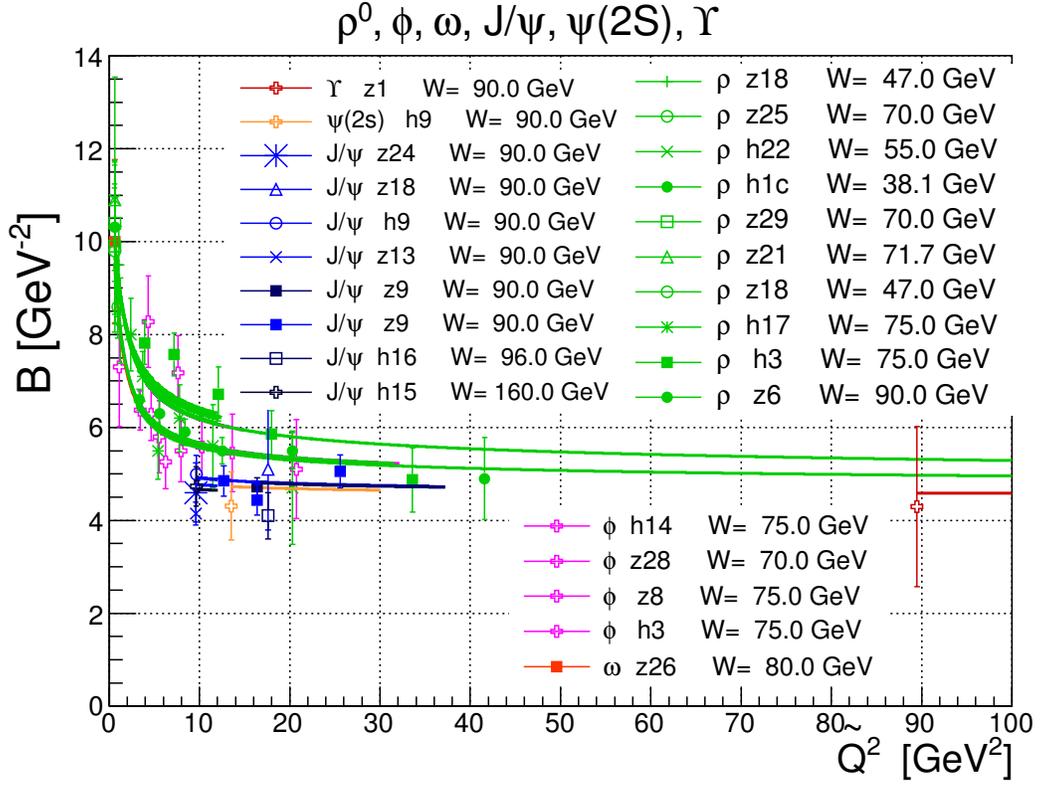}
\caption{ \label{fig:B(Q2)} Experimental data on the slope $B$ as function of $\widetilde{Q^2}$ for $\rho^0, \phi, J/\psi$, $\varUpsilon$ and $\Psi$(2S), and~our theoretical predictions from Equation~(\ref{eq:B(h+s)}).}
%For a compilations of the slopes for different $|t|$ bins, see Figure~\ref{fig:B(Q2)_in_t}.}
\end{figure}

Next, by~using Equation~\eqref{eq:dcsdt(h+s)} with the values of the parameters from Table~\ref{tab:fit1(s+h)} and the formula
\begin{equation}\label{eq:B(h+s)}
B(Q^2,W,t)=\frac{d}{dt}\ln{\frac{d\sigma_{el}}{dt}},
\end{equation}
one calculates the forward slopes and compares them with the experimental data on VMP, including those for the $\Psi$(2S) production.  A~compilation of all results is presented in  Figure~\ref{fig:B(Q2)}.

The number of fitted parameters of the two-component pomeron model (Equation~\eqref{eq:f_i})  is 12 (Table~\ref{tab:fit1(s+h)}), with~five additional normalization factors%
for six vector particle productions ($\rho^0$, $\phi$, $\omega$, $J/\psi$, $\varUpsilon$ and $\gamma$). By~fixing the pomeron trajectories $\alpha_s(t)$ and $\alpha_h(t)$ (see Table~\ref{tab:fixed_trajec}), the~number of free parameters reduces to $8$.

%\end{document}

%\newpage
%### Figure~5 ######
\unskip

%### Figure~7 ######

%\end{document}

\unskip

%### Figure~9 ######
\unskip

%### Figure~10 ######
%\newpage

%### Figure~11 ######
\unskip

%### Figure~20 ######

% Table~4
\begin{table}[h]
\centering
\caption{ Parameters of the two-component pomeron model (Equations~(\ref{eq:dcsdt(h+s)})
and (\ref{eq:cs(h+s)})) fitted to the combined VMP and DVCS data, with~fixed    parameters of the pomeron trajectories $\alpha_{s}(t)=1.08+0.25t$ and $\alpha_{h}(t)=1.20+0.01t$.}
\label{tab:fixed_trajec}
%$\tilde\chi^2/Ndf=1.00$.}\label{tab:fit2(s+h)}
\begin{tabular}{cc c c c c c c}
\toprule
&\boldmath{$A_{0s,h}$} $\left[\frac{\sqrt{\text{\textbf{nb}}}}{\text{\textbf{GeV}}}\right]$
&\boldmath{$\widetilde{Q^2_{s,h}}$}  $\left[\text{\textbf{GeV}}^\textbf{2}\right]$
&\boldmath{$n_{s,h}$} & \boldmath{$\alpha_{0\,s,h}$}
&\boldmath{$\alpha'_{s,h}$} $\left[\frac{\textbf{1}}{\text{\textbf{GeV}}^\textbf{2}}\right]$
&\boldmath{$b_{s,h}$}  $\left[\frac{\textbf{1}}{\text{\textbf{GeV}}^\textbf{2}}\right]$\\ \midrule
soft&807 $\pm$ 1107&0.46 $\pm$ 0.70&1.79 $\pm$ 0.79&1.08&0.25&3.41 $\pm$ 2.48\\
hard& 47.9 $\pm$  46.9&1.30 $\pm$ 1.12&1.33 $\pm$ 0.26&1.20&0.01&2.15 $\pm$ 1.14\\
\bottomrule
\end{tabular}
\end{table}

{More results and discussion on the fitting details can be found in Refs.~\cite{Salii1, Salii2}}.

%\newpage
\section{Balancing between ``Soft'' and ``Hard''~Dynamics}\label{sec:Balance}
In this section we illustrate
%in various ways
the important and delicate interplay between the ``soft'' and ``hard'' components of our unique pomeron.
Since the amplitude consists of two parts, according to the definition ~(\ref{two-term-amp}), it can be written as
\begin{equation}
A(Q^2,s,t)=A_s(Q^2,s,t)+A_h(Q^2,s,t).
\label{Ampl-2}
\end{equation}

As a consequence, the~differential and elastic cross sections contain also an interference term between ``soft'' and ``hard'' parts, so that they read
\begin{equation}
\frac{d\sigma_{el}}{dt}=\frac{d\sigma_{s,el}}{dt}+\frac{d\sigma_{h,el}}{dt}+\frac{d\sigma_{interf,el}}{dt}
\label{dsigma_2}
\end{equation}
and
\begin{equation}
\sigma_{el}=\sigma_{s,el}+\sigma_{h,el}+\sigma_{interf,el},
\label{sigma_2}
\end{equation}
according to Equations~\eqref{eq:dcsdt(h+s)} and \eqref{eq:cs(h+s)}, respectively.

%### Figure~23 ######

%\newpage

Given Equations~(\ref{dsigma_2}) and (\ref{sigma_2}), we can define the following ratios for each component:
\begin{equation}
R_i(\widetilde{Q^2}, W, t)=\frac{ \frac{d\sigma_{i,el}}{dt} }{ \frac{d\sigma_{el}}{dt} }
\label{ratio_dsigma}
\end{equation}
and
\begin{equation}
R_i(\widetilde{Q^2}, W)=\frac{\sigma_{i,el}}{\sigma_{el}},
\label{ratio_sigma}
\end{equation}
where $i$ stands for $\{s, h, interf\}$.

%(see Figure~\ref{fig:Rsh_surf}, where ; and as $ for integrated cross sections (see  (right)).

\begin{figure}[h]
\centering
\includegraphics[trim = 0mm 0mm 0mm 0cm,clip, scale=0.42]{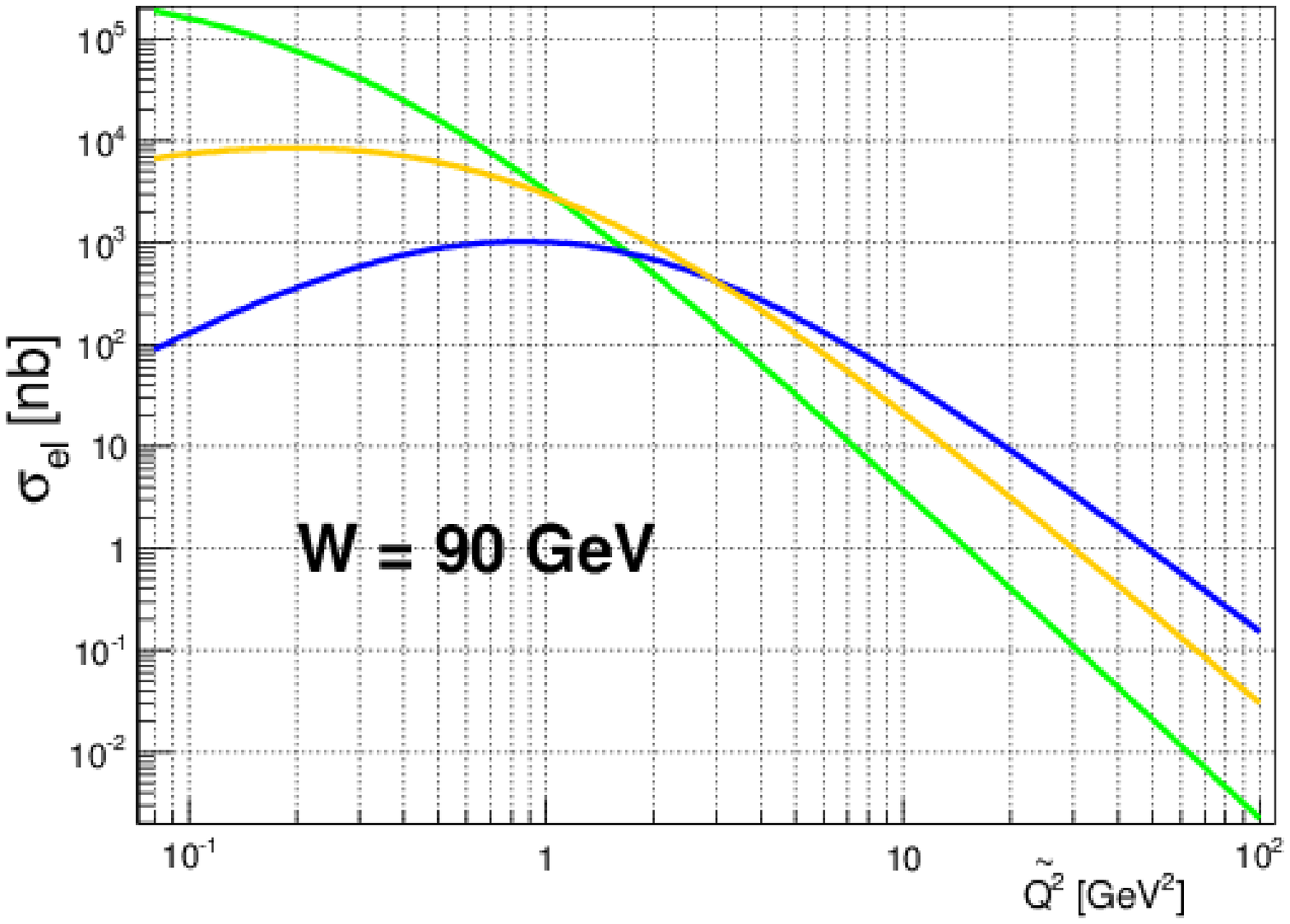}
\includegraphics[trim = 0mm 0mm 0mm 0cm,clip, scale=0.43]{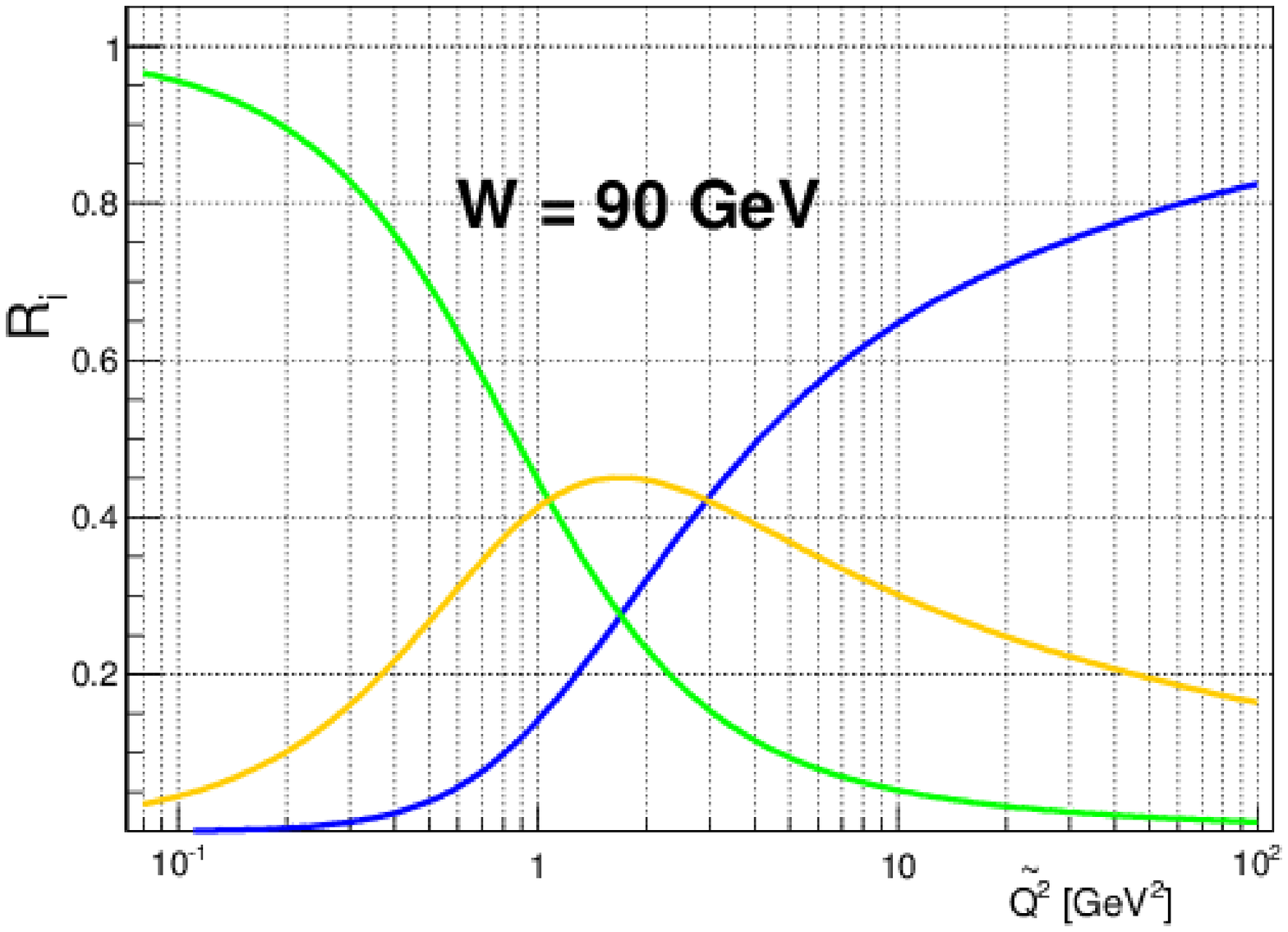}
\caption{\label{fig:Rsh} Interplay between soft (green line), hard (blue line) and interference (yellow line) components of the cross section $\sigma_{i,el}$ (left plot) and $R_i(\widetilde{Q^2}, t)$ (right plot) as functions of $\widetilde{Q^2}$, \mbox{for~$W=70$~GeV}.}
\end{figure}

\begin{figure}[h]
\centering
\includegraphics[trim = 2mm 1mm 0mm 0cm,clip, scale=0.38]{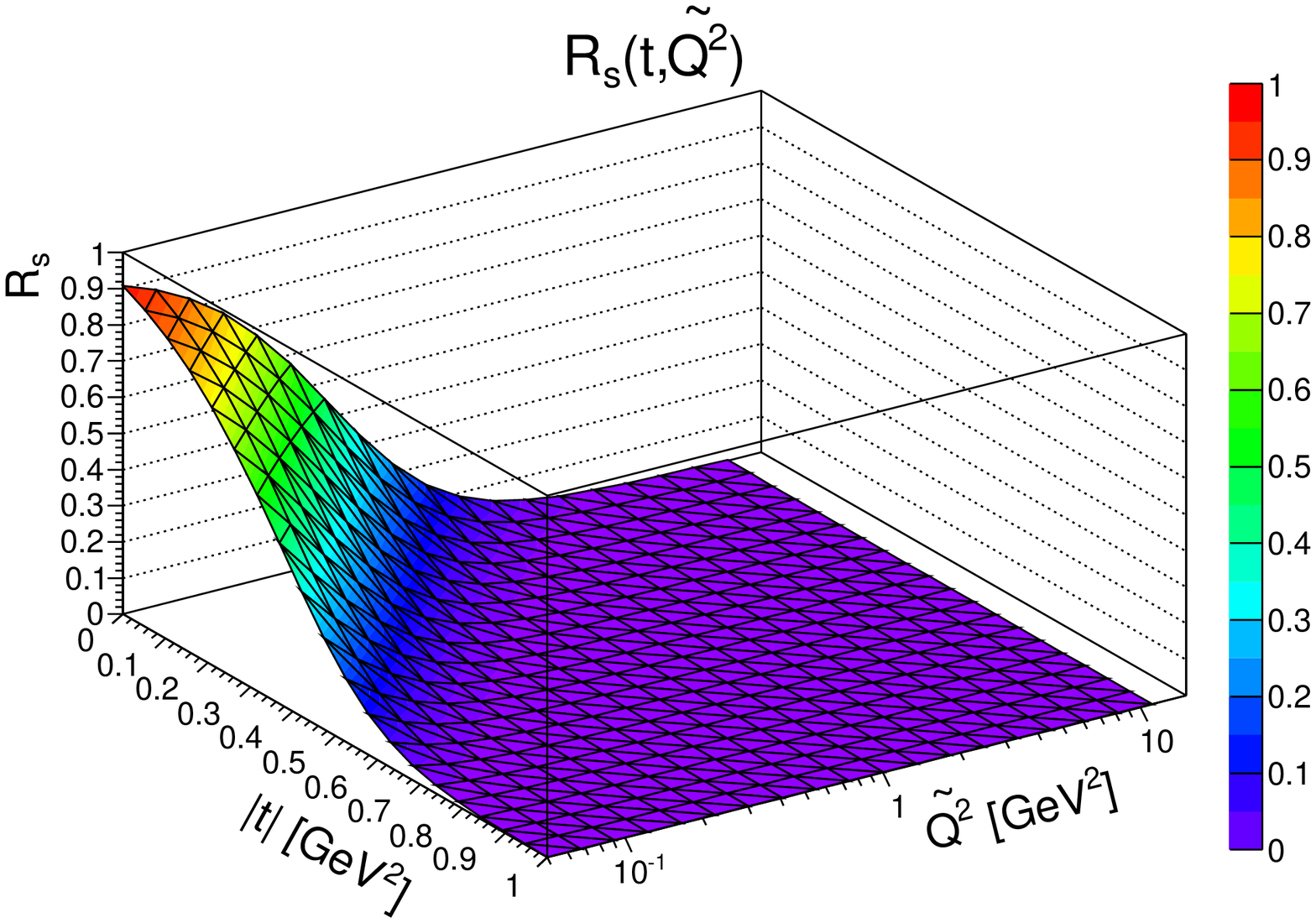}
\includegraphics[trim = 0mm -5mm 0mm 13mm,clip, scale=0.34]{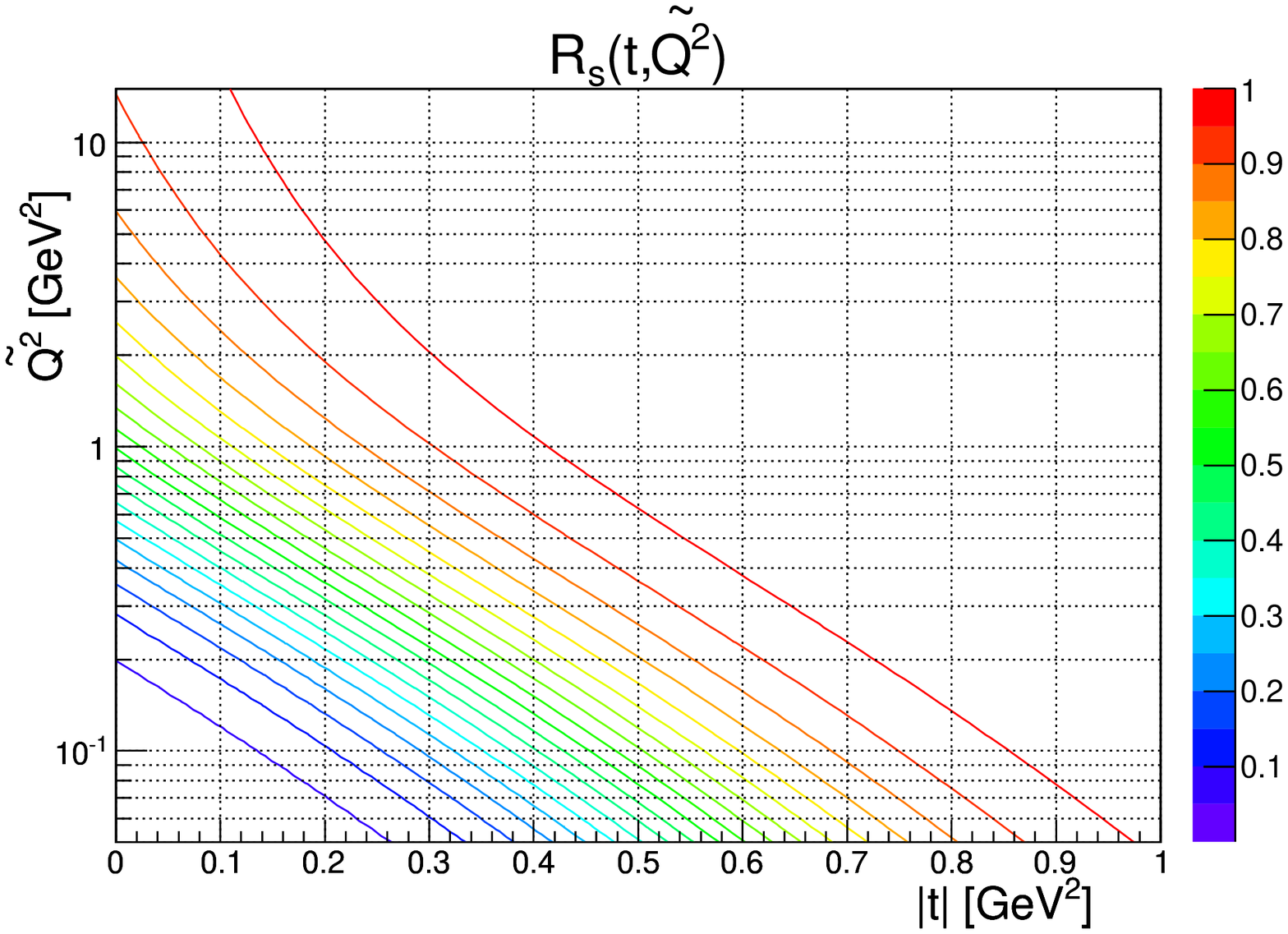}\\
\includegraphics[trim = 3mm 1mm 0mm 0cm,clip, scale=0.38]{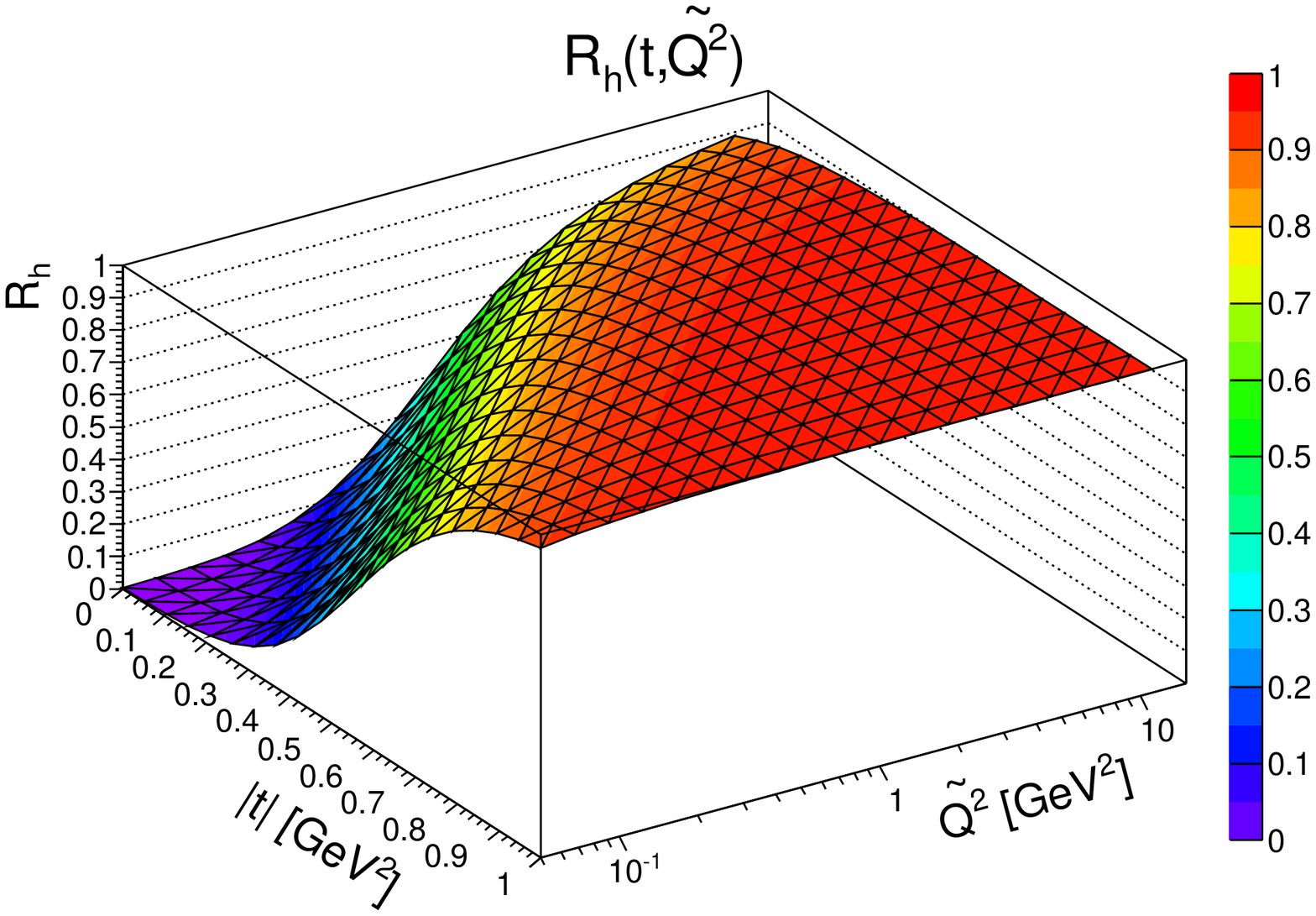}
\includegraphics[trim = 0mm -5mm 0mm 13mm,clip, scale=0.34]{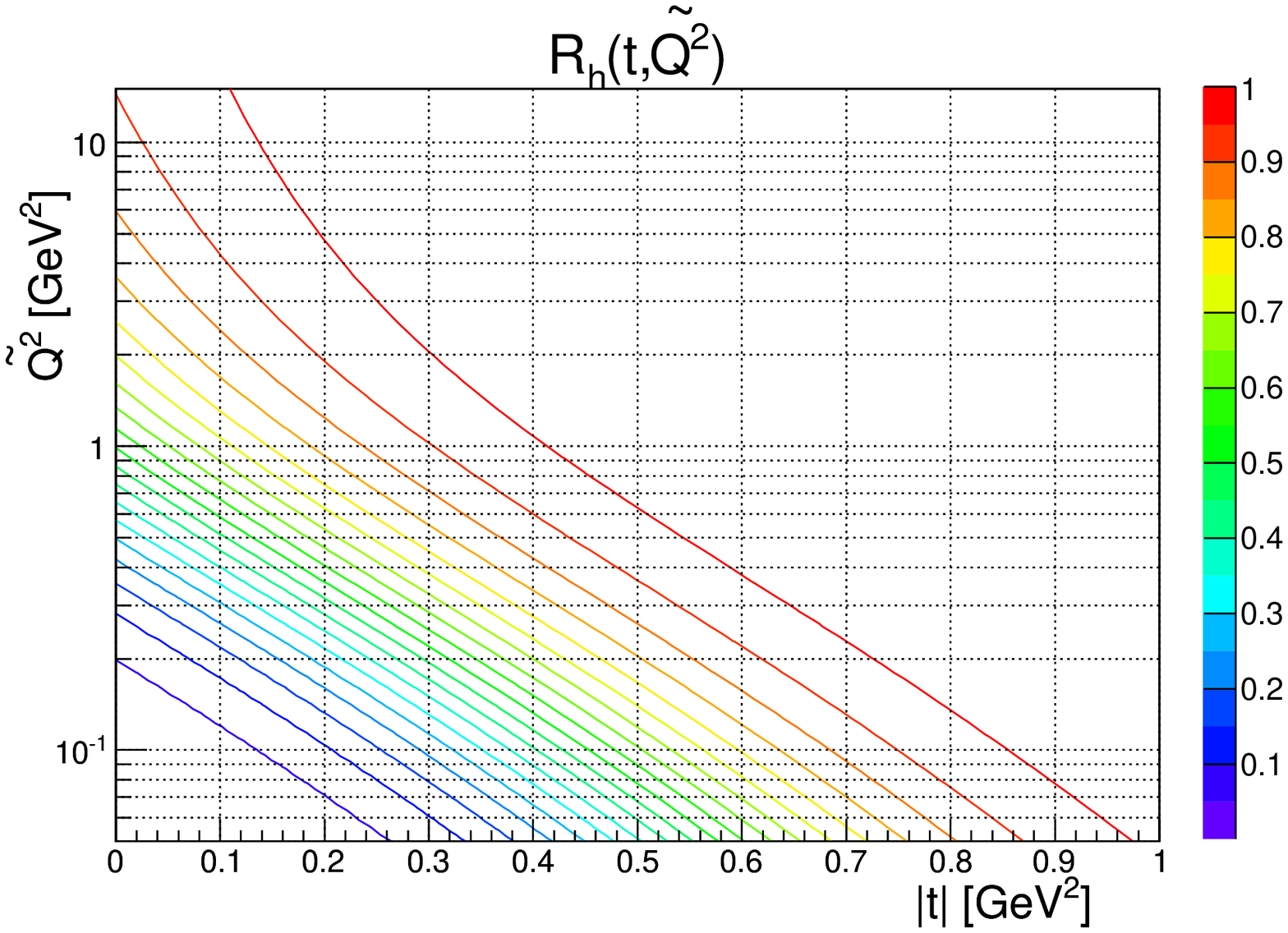}\\
\includegraphics[trim = 2mm 1mm 0mm 0cm,clip, scale=0.38]{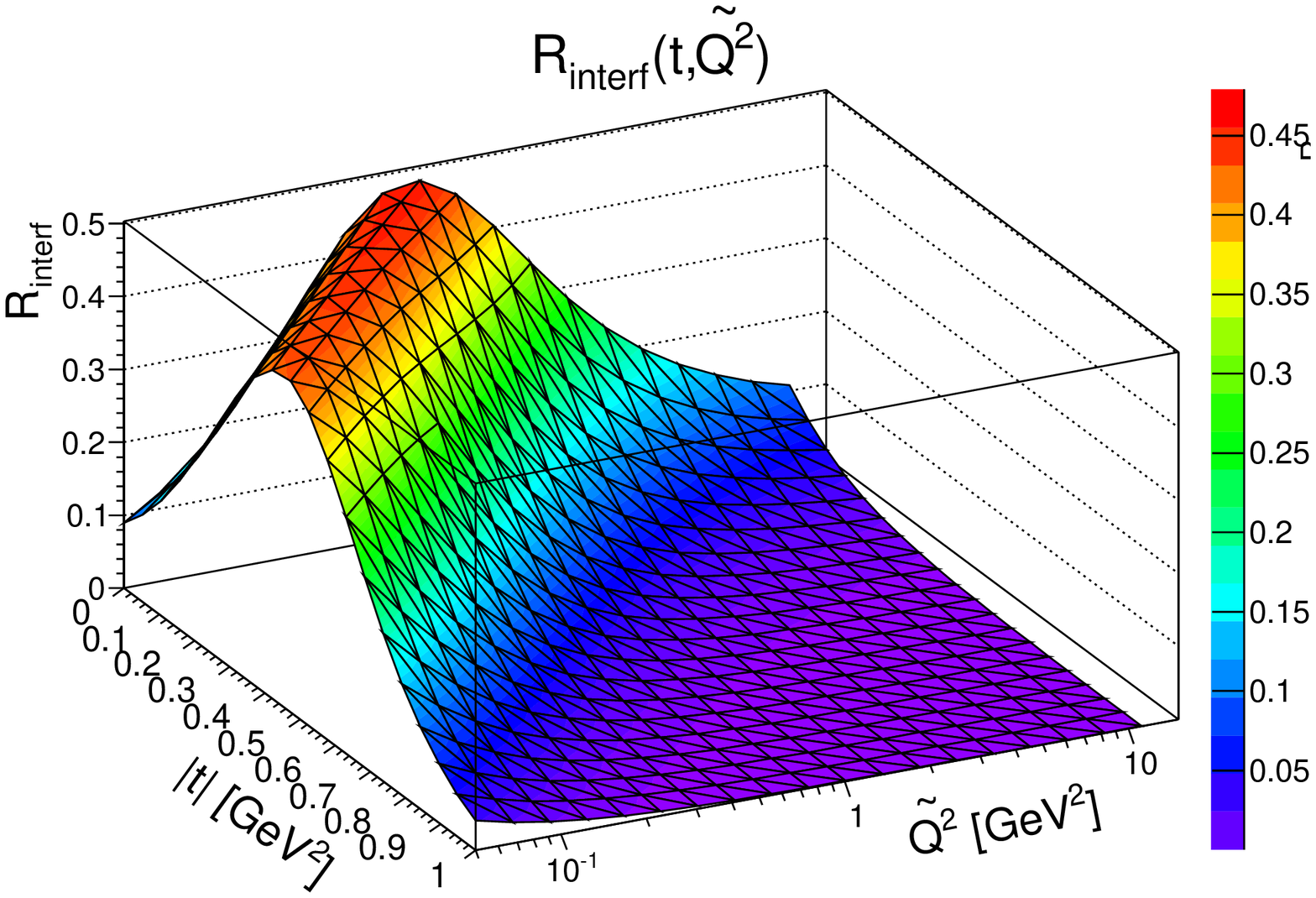}
\includegraphics[trim = 0mm -5mm 0mm 13mm,clip, scale=0.34]{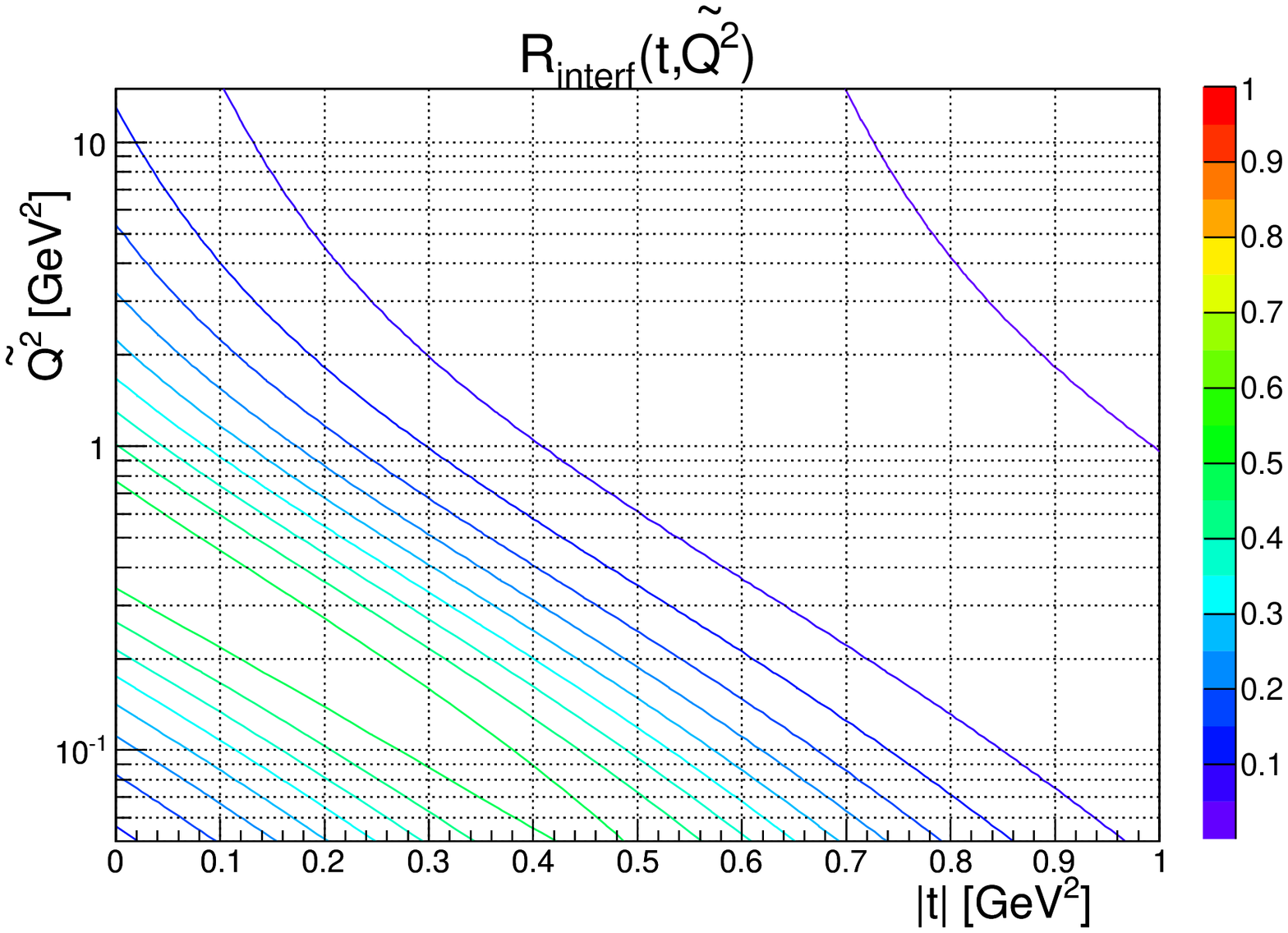}
\caption{\label{fig:Rsh_surf} \textbf{Left column}: soft (\textbf{upper surface}), hard (\textbf{middle surface}) and interference (\textbf{bottom surface}) components of the ratio $R_i(\widetilde {Q^2}, W, t)$
%(see Equation~(\ref{ratio_dsigma}))
are shown  as functions of $\widetilde{Q^2}$ and $t$, for~$W=70$~GeV. \textbf{Right column}: some representative curves of the surfaces projected on the ($t, \widetilde {Q^2}$  ) plane.}
\end{figure}

Figure~\ref{fig:Rsh} shows the interplay between the components for both $\sigma_{i,el}$ and $R_i(\widetilde{Q^2}, t)$, as~functions of $\widetilde {Q^2}$, for~$W$ = 70  GeV.
In Figure~\ref{fig:Rsh_surf} both plots show that not only $\widetilde{Q^2}$ is the parameter defining softness or hardness of the process, but~such is also the  combination of $\widetilde{Q^2}$ and $t$, similar to the variable $z=t-Q^2$ introduced in Ref.~\cite{Capua}.
On the whole, it can be seen from the plots that the soft component dominates in the region of low $\widetilde{Q^2}$ and low $|t|$, while the hard component dominates high $\widetilde{Q^2}$ and high $|t|$. %We notice that the behaviour of the components is similar to that of Figure~\ref{fig:Rsh_surf}.

In other words, the~variables $t$ and $Q^2$ have much in common: both refer to squared momentum transfer and have the meaning of ``hardness''. The~present model provides a valuable laboratory to study this important~issue.

Hadron-induced reactions, discussed in Section~\ref{Sec:Hadron} differ from those induced by photons at least in two aspects. First, hadrons are on the mass shell and hence the relevant processes are typically ``soft''. Secondly, the~mass of incoming hadrons is positive, while the virtual photon has negative squared ``mass''. Our attempt to include hadron-hadron scattering into the analysis with our model has the following motivations: (a)
by vector meson dominance (VMD) the photon behaves partly as a meson, therefore meson--baryon (and more generally, hadron-hadron) scattering has much in common with photon-induced reactions. Deviations from VMD may be accounted for the proper $Q^2$ dependence of the amplitude (as we do hope is in our case!); (b) connection between
space- and time-like reactions;  (c)~according to some claims the highest-energy (LHC) proton--proton scattering data indicate the need for a ``hard'' component in the pomeron (to anticipate, our fits do not found support the need of any noticeable ``hard'' component in $pp$ scattering).

We do not intend to perform here
a high-quality fit to the $pp$ data; that would be impossible without the inclusion of subleading contributions and/or the~odderon.

The $pp$ scattering amplitude is written in the form similar to the amplitude  (\ref{eq:Amplitude_hs}) for VMP or DVCS, the~only difference being that the normalization factor is constant since the $pp$ scattering amplitude does not depend on $Q^2$:
%\paragraph{For two-component Pomeron.}
%The amplitude for $pp$-scattering was constructed in the form of Eq.~\eqref{eq:Amplitude_hs} where $Q^2$-dependent factors $H_{s,h}$ were substituted by constants, since the $pp$ amplitude does not depend on $Q^2$.
\begin{equation}\label{eq:Amplitude2_pp}
A^{pp}(s,t)=
A^{pp}_s\, e^{-i\frac{\pi}{2}\alpha_s(t)} \left(\frac{s}{s_{0}}\right)^{\alpha_s(t)} e^{b_st}
+A^{pp}_h\, e^{-i\frac{\pi}{2}\alpha_h(t)} \left(\frac{s}{s_{0}}\right)^{\alpha_h(t)} e^{b_ht}.
\end{equation}

%Consequently, the differential and integrated elastic cross sections take the form of Equations~(\ref{eq:dcsdt(h+s)})
%and (\ref{eq:cs(h+s)}), with only the substitution of $H_s$ and $H_h$ with $A^{pp}_s$ and $A^{pp}_h$, respectively.

We fix the parameters of pomeron trajectory at
$$\alpha_{s}(t)=1.084+0.35t,\qquad \alpha_{h}(t)=1.30+0.10t.$$

With these trajectories the total cross section
\begin{equation}\label{eq:cstot_2}
\sigma_{tot}=\frac{4\pi}{s} Im\;A(s,{t=0})
\end{equation}
was found compatible with the LHC data. From~the comparison of Equation~(\ref{eq:cstot_2}) to the LHC data we~get
$$A^{pp}_s=-1.73 \text{\,mb}\cdot\text{GeV}^2,\,\,\quad\quad
A^{pp}_h=-0.0012 \text{\,mb}\cdot\text{GeV}^2.$$

We conclude that, while the data on total cross section are compatible with a small ``hard" admixture in the amplitude, the~slope parameter with a hard component included seems to manifest a wrong tendency, by~slowing down with increasing energy, while the TOTEM measurements  show the~opposite.

The resulting fits are reasonable, despite the following open problems:

\begin{itemize}
\item%4
sub-leading Regge contributions must be included in any extension of the model to lower energies (below $30$ GeV);
%We intend to do so in a future study.

\item%5
the $\widetilde{Q^2}$ dependence of the scattering amplitude, introduced empirically has to be compared with the results of unitarization and/or QCD~evolution.

\item%6
as seen from Section~\ref{sec:Balance}, the~``soft'' component of the pomeron dominates in the region of small $t$ and small $\widetilde{Q^2}$. Hence, any parameter responsible for the ``softness'' and/or ``hardness'' of processes, should be a combination of $t$ and $Q^2$. A~simple solution was suggested in Ref.~\cite{Capua} with the introduction of the variable $z=t-Q^2$. The~interplay of these two variables remains  an important open problem that requires further investigation.
\end{itemize}

The extension of our formalism to hadronic reactions ($pp$ scattering) shows that
%the behavior of total cross sections allows  for the presence of a small ``hard" component.
%But,the rise with energy of the forward slope slows down, while the TOTEM point requires opposite.
the available data can be will described by %only one --
a single---soft---component.

\section{Spin of the Proton in Terms of Its~Constituents}\label{Sec:Crisis}

{
The origin of spin in the nucleon is a rich subject deserving a dedicated presentation. Here I only briefly sketch some recent developments in the field. A~complete and comprehensive treatment of the subject can be found e.g.,~in Refs.~\cite{Leader0, Rev2}.

In the simplest case, if~only valence quarks contributed, the~nucleon spin would be simple a sum
\begin{equation} \label{spin}
J=\frac{\Delta\Sigma}{2}+L_q=\frac{1}{2},
\end{equation}
where $\Delta\Sigma$ is the quark spin contribution and $L_q$ is the orbital angular momentum (OAM) contribution. Partition between the quark spin and the OAM for a long time were subject of debates generally called “spin crisis”. Estimates of shares varied in a wide range. Among~the reasons of disagreement and contradictions was the role of gluons, and~sea as compared to that of the valence quarks.
The~phenomenon known as “spin crisis” arose from an EMC experiment at CERN~\cite{EMC} quoting $\Sigma\approx 0$ (with large uncertainty), thus allegedly contradicting the naive quark model. Recent COMPAS, HERMES and Jlab measurements are consistent with a value of  $\Sigma\approx 0.3.$ Actually, $J$ must obey conservation of the total angular momentum known as Jaffe-{Manohar sum rule}~\cite{Jaffe_Manohar},
%mdpi: There is no {Jaffe-Manohar} find in this paper, please add.
\begin{equation}\label{J-M}
J=\Delta\Sigma(Q^2)+L_q(Q^2)+\Delta(Q^2)+L_q(Q^2)=\frac{1}{2}.
\end{equation}

An important task in the context of the above some rule is the separation of sea and valence quarks, indistinguishable in DIS observables. In addition, as~ {noted in Ref.}~\cite{BMa}, strange and anti-strange sea quarks can contribute differently in to the nucleon~spin.
%%mdpi: There is no {B=Ma} find in this paper, please add.

Paper~\cite{Rev2} discusses also how DIS data on proton, neutron and deuteron targets can be used to separate the contributions from different quark polarizations assuming validity of $SU(3)_f$.

That paper contains many more ideas, among~which is the use of parton--hadron duality of Bloom and Gilman~\cite{B-G}---another effective tool to be used in studies of spin effects and a possible key to understand better problems related to quark confinement.}

%\begin{comment}
Before the advent of QCD, a~nucleon (N) was visualized as a bound state  of 3 massive quarks (Q) ( $ M_Q \approx  M_N/3 $)  lying in some kind of potential.   In~the simplest non-relativistic case, for~an s-state the constituent quarks have no orbital angular momentum (OAM) and one has, for~the nucleon at rest, say~polarized in the positive  Z-direction
\begin{equation}
\label{naive-spin} 1/2 =   S_z^{N}= \sum_Q S_z^Q.\end{equation}

With relativistic corrections, the~bottom  components of the quark Dirac spinors contain OAM and Equation~(\ref{naive-spin}) is modified to
\begin{equation}
\label{rel-naive} \sum_Q S_z^Q \approx 0.3 \end{equation}

In the {naive} parton model, for~a fast moving proton with helicity $+1/2$,
\begin{equation}
\label{simple} a_0= \Delta \Sigma \equiv (\Delta u + \Delta\overline{u}) + (\Delta d + \Delta\overline{d}) + (\Delta s + \Delta\overline{s})
\end{equation}
where the $\Delta q $, $\Delta \bar{q}$ are the first moments of the polarized quark--parton helicity densities. Hence, bearing in mind that $\Delta q = q_{+} -q_{-} $, where $\pm $ corresponds to the number densities with spin along or opposite to the proton's momentum, one obtains  in the naive parton model
\begin{equation}
a_0 = \Delta \Sigma =2 \left[\sum_{q} \langle  S_z^{q} \rangle + \sum_{\bar{q}} \langle  S_z^{\bar{q}} \rangle\right] \end{equation}
and if there is no other source of angular momentum one expects
\begin{equation}
\label{wrong} \left[\sum_{q} \langle  S_z^{q} \rangle + \sum_{\bar{q}} \langle  S_z^{\bar{q}} \rangle\right]= S_z^{\textrm{proton}}= 1/2 \end{equation}
implying, naively,
$$\label{nivespin} a_0 = 1. $$

After the famous EMC experiments revealed that only a small fraction
of the nucleon spin is due to quark spins,
there has been great
interest in `solving the spin puzzle', i.e.,~in decomposing the
nucleon spin into contributions from quark/gluon spin and
orbital degrees of~freedom.

\textls[-20]{The EMC experiment gave $a_0\approx 0 $ and later experiments confirmed that $a_0 \ll 1 $, giving rise to the spin crisis in the (\emph{naive}) parton model. However,  Equation~(\ref{nivespin}) cannot possibly be true because the right hand side is a fixed number, whereas the left hand side is, {beyond the naive level}, equal to $a_0(Q^2)$, i.e.,~a~function of $Q^2$! Thus failure of Equation~(\ref{nivespin}) to hold cannot be used to infer that there is  crisis. It is obvious that a correct relation between the spin of a  nucleon and the angular momentum of its constituents should include their orbital angular momentum and should also include a contribution from the~gluons.}

%\end{comment}

\subsection*{Ji's and J-M's~Decompositions}
Finally, we briefly mention two sum rules visualising in different albeit complementary ways
the spin decomposition in the~nucleon.

The Ji decomposition~\cite{Ji:1996ek}
\begin{equation}
\frac{1}{2}=\frac{1}{2}\sum_q\Delta q + \sum_q {L}_q^z+
J_g^z
\label{eq:JJi}
\end{equation}
appears to be very useful,
as not only the quark spin contributions $\Delta q$ but also
the quark total angular momenta $J_q \equiv \frac{1}{2}\Delta q +
{ L}_q^z$ (and by subtracting the spin piece also the
the quark orbital angular momenta $L_q^z$) entering this decomposition
can be accessed experimentally, through generalized parton
distributions (GPDs).
The terms in (\ref{eq:JJi}) are defined as expectation
values of the corresponding terms in the angular momentum tensor
\begin{equation}
M^{0xy}= \sum_q \frac{1}{2}q^\dagger \Sigma^zq +
\sum_q q^\dagger \left({\vec r} \times i{\vec D}
\right)^zq
+
\left[{\vec r} \times \left({\vec E} \times {\vec B}\right)\right]^z
\label{M012}
\end{equation}
in a nucleon state with zero momentum. Here
$i{\vec D}=i{\vec \partial}-g{\vec A}$ is the gauge-covariant
derivative.

Jaffe and Manohar proposed~\cite{Jaffe_Manohar} an alternative decomposition of the
nucleon spin, which does have a partonic interpretation
\begin{equation}
\frac{1}{2}=\frac{1}{2}\sum_q\Delta q + \sum_q {\cal L}_q^z+
\frac{1}{2}\Delta G + {\cal L}_g^z,
\label{eq:JJM}
\end{equation} whose terms are defined as matrix elements of the corresponding
terms in the $+12$ component of the angular momentum tensor
\begin{equation}
M^{+12} = \frac{1}{2}\sum_q q^\dagger_+ \gamma_5 q_+ +
\sum_q q^\dagger_+\left({\vec r}\times i{\vec \partial}
\right)^z q_+
+ \varepsilon^{+-ij}\mbox{Tr}F^{+i}A^j
+ 2 \mbox{Tr} F^{+j}\left({\vec r}\times i{\vec \partial}
\right)^z A^j.
\label{M+12}
\end{equation}

\begin{figure}[h]
\centering
\includegraphics[scale=0.73]{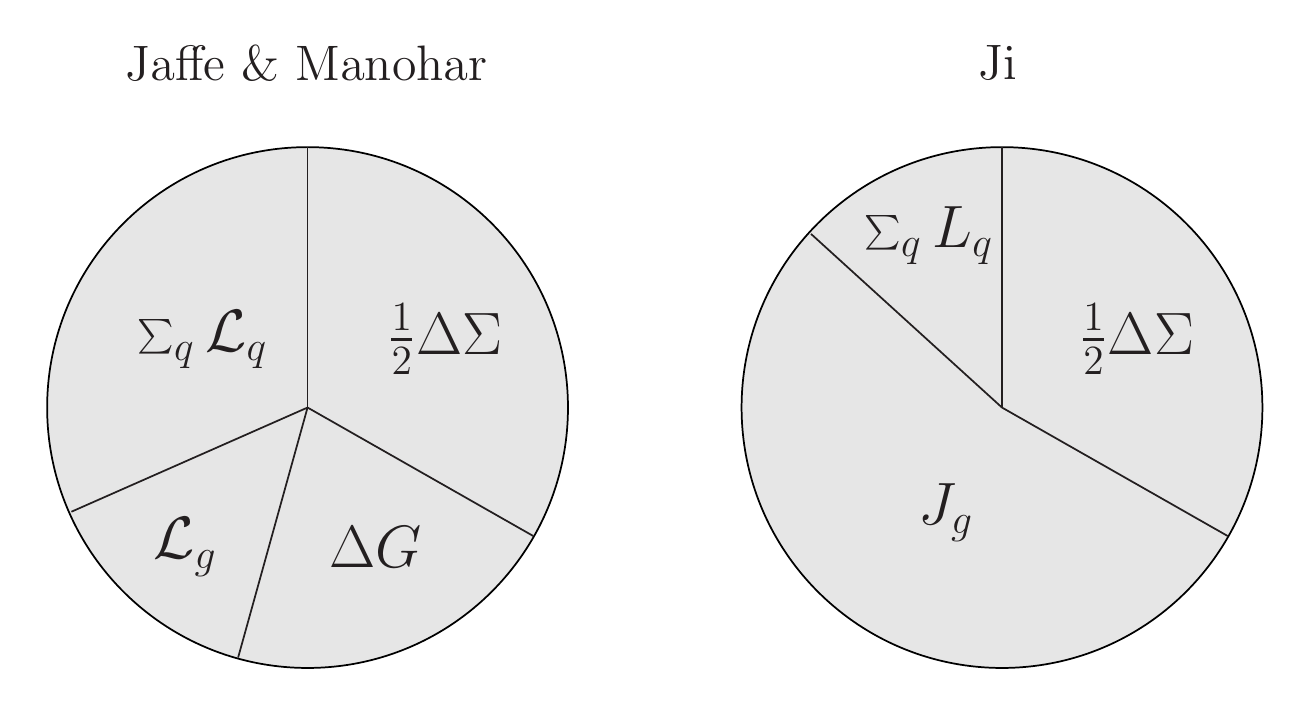}
\caption{{Origin of the proton} spin according to sum rules of Jaffe, and Monahor and~Ji.}
%mdpi: Please complete the picture of Figure 17. I tried to take a screenshot to fix it myself, but the part on the right is always missin.
\label{fig:pizza}
\end{figure}

In Figure~\ref{fig:pizza} the share of various components to the nucleon spin, according to the two (Ji and Jaffe-Manohar) distribtutions is~shown.

%\newpage
%### Figure~7 ######

%\subsection{No "spin crisis"}\label{secI}

%\subsection{The naive interpretation of the EMC results}
%\subsection{Conclusion}
What appeared to be spin crisis in the parton model, over~three decades years ago, was a consequence of a misinterpretation of the results of
the famous European Muon Collaboration experiment on polarized deep inelastic~scattering.

\section{Conclusions}
Interest in spin physics and polarization has moved from hadron-hadron to lepton-hadron processes. This is mainly due to the fact that in proton--proton and proton--antiproton scattering spectacular spin effects were/are not expected at highest energies (ISR, Fermilab, BNL or LHC), except~maybe for the dip-bump structure at the diffraction cone. Various aspects of polarization in inelastic hadronic reactions  reactions were treated in simple intuitive models in Refs.~\cite{Strum}.

Instead, deeply virtual Compton scattering (DVCS), with~spin degrees of freedom is now at the focus of interest as a source of information on generalized parton distributions (GPDs).

In both cases (hadron-hadron and lepton-hadron) diffraction and the pomeron play a crucial role. Therefore much space was dedicated to the theoretical problems of diffractive scattering, properties of the vacuum Regge trajectory (pomeron), and~confinement of quarks and~gluons.

In the near future, research programs at the  Electron-Ion Colliders (EICs) will play an important role in understanding gluon GPDs. Apart from providing
constraints on the total quark/gluon contributions to the proton spin, the~GPD’s will provide important information on the
nucleon tomography, for~example, the~3D imaging of partons inside the proton. Together with the gravitational form factors extracted
from the DVCS, this will deepen our understanding of the nucleon spin structure in return.
The EIC may shed light on the quark/gluon orbital angular momentum (OAM) directly through various hard diffractive processes.  Pioneer experimental effort to constrain the gravitation form factor from DVCS experiment at JLab has been carried out in Ref.~\cite{Grav}.
The quark and gluon helicity contributions to the proton spin with the unique coverage in both $x$ and $Q^2$ at EIC will provide the most stringent constraints on $\Delta \Sigma$ and $\Delta G$ \cite{EIC1, EIC2}.

Last but not least, Regge trajectories Section \ref{Subs:Trajectory}, relating uniquely the spin and mass of particles, are building blocks of the theory, containing the basic information on the dynamics. By~crossing symmetry, unitarity and duality they relate various aspects of high-energy strong interaction~dynamics.

\vspace{6pt}
\textbf{{This research received no external funding}}
%mdpi: Please add: This research received no external funding or This research was funded by [name of funder] grant number [xxx] And The APC was funded by [XXX].
\acknowledgments{During many years of my professional activity I enjoyed and profited from the collaboration with Salvatore Fazio, Roberto Fiore, Francesco Paccanoni, Alessandro Papa, Enrico Predazzi, Rainer
Schicker, Istv{\'a}n Szanyi and Andrii Salii as well from discussions with  Victor Fadin, Sergey Troshin and Oleg Selyugin, whom I thank very much. I acknowledge the intelligent and useful remarks by the Referee that helped me to improve this presentation.
The work was supported by the National Academy of Sciences of Ukraine, grant N 012r100935, ``Fundamental properties of   matter in relativistic collisions of nuclei and in the early~Universe''.}
\textbf{{The author declares no conflict of interest.} %mdpi: Please disclose any conflicts of interest, or add “The authors declare no conflicts of interest.
}

\center{\textbf{{References}}}
%mdpi: References 50-60 are not mentioned in the main-text, please add. And add the title of references which are no title.

%%%%%%%%%%%%%%%%%%%%%%%%%%%%%%%%%%%%%%%%%%%%%%%%%%%%%%%%%%%%%%%%%
\end{document}